\documentstyle[12pt,preprint]{aastex}  

\begin{document}

\title{Light Curves of Dwarf Plutonian Planets and other Large Kuiper Belt Objects: 
Their Rotations, Phase Functions and Absolute Magnitudes}  
\author{Scott S. Sheppard}    
\affil{Department of  Terrestrial Magnetism, Carnegie Institution of Washington, \\
5241 Broad Branch Rd. NW, Washington, DC 20015 \\ sheppard@dtm.ciw.edu}

\begin{abstract}  

I report new time-resolved light curves and determine the rotations
and phase functions of several large Kuiper Belt objects, which
includes the dwarf planet Eris (2003 UB$_{313}$).  Three of the new
sample of ten Trans-Neptunian objects display obvious short-term
periodic light curves. (120348) 2004 TY$_{364}$ shows a light curve
which if double-peaked has a period of $11.70\pm0.01$ hours and a
peak-to-peak amplitude of $0.22\pm 0.02$ magnitudes.  (84922) 2003
VS$_{2}$ has a well defined double-peaked light curve of $7.41\pm0.02$
hours with a $0.21\pm 0.02$ magnitude range.  (126154) 2001 YH$_{140}$
shows variability of $0.21\pm 0.04$ magnitudes with a possible
$13.25\pm0.2$ hour single-peaked period.  The seven new KBOs in the
sample which show no discernible variations within the uncertainties
on short rotational time scales are 2001 UQ$_{18}$, (55565) 2002
AW$_{197}$, (119979) 2002 WC$_{19}$, (120132) 2003 FY$_{128}$,
(136108) Eris 2003 UB$_{313}$, (90482) Orcus 2004 DW, and (90568) 2004
GV$_{9}$.  Four of the ten newly sampled Kuiper Belt objects were
observed over a significant range of phase angles to determine their
phase functions and absolute magnitudes.  The three medium to large
sized Kuiper Belt objects 2004 TY$_{364}$, Orcus and 2004 GV$_{9}$
show fairly steep linear phase curves ($\sim$ 0.18 to 0.26 mags per
degree) between phase angles of 0.1 and 1.5 degrees.  This is
consistent with previous measurements obtained for moderately sized
Kuiper Belt objects.  The extremely large dwarf planet Eris (2003
UB$_{313}$) shows a shallower phase curve ($0.09\pm 0.03$ mags per
degree) which is more similar to the other known dwarf planet Pluto.
It appears the surface properties of the largest dwarf planets in the
Kuiper Belt maybe different than the smaller Kuiper Belt objects.
This may have to do with the larger objects ability to hold more
volatile ices as well as sustain atmospheres.  Finally, it is found
that the absolute magnitudes obtained using the phase slopes found for
individual objects are several tenths of magnitudes different than
that given by the Minor Planet Center.

\end{abstract}


\section{Introduction}

To date only about $1\%$ of the Trans-Neptunian objects (TNOs) are
known of the nearly one hundred thousand expected larger than about 50
km in radius just beyond Neptune's orbit (Trujillo et al. 2001).  The
majority of the largest Kuiper Belt objects (KBOs) now being called
dwarf Plutonian planets (radii $>$ 400 km) have only recently been
discovered in the last few years (Brown et al. 2005).  The large self
gravity of the dwarf planets will allow them to be near Hydrostatic
equilibrium, have possible tenuous atmospheres, retain extremely
volatile ices such as Methane and are likely to be differentiated.
Thus the surfaces as well as the interior physical characteristics of
the largest TNOs may be significantly different than the smaller TNOs.

The largest TNOs have not been observed to have any remarkable
differences from the smaller TNOs in optical and near infrared broad
band color measurements (Doressoundiram et al. 2005; Barucci et
al. 2005).  But near infrared spectra has shown that only the three
largest TNOs (Pluto, Eris (2003 UB$_{313}$) and (136472) 2005
FY$_{9}$) have obvious Methane on their surfaces while slightly
smaller objects are either spectrally featureless or have strong water
ice signatures (Brown et al. 2005; Licandro et al. 2006).  In addition
to the Near infrared spectra differences, the albedos of the larger
objects appear to be predominately higher than those for the smaller
objects (Cruikshank et al. 2005; Bertoldi et al. 2006; Brown et
al. 2006).  A final indication that the larger objects are indeed
different is that the shapes of the largest KBOs seem to signify they
are more likely to be in hydrostatic equilibrium than that for the
smaller KBOs (Sheppard and Jewitt 2002; Trilling and Bernstein 2006;
Lacerda and Luu 2006).

The Kuiper Belt has been dynamically and collisionally altered
throughout the age of the solar system.  The largest KBOs should have
rotations that have been little influenced since the sculpting of the
primordial Kuiper Belt. This is not the case for the smaller KBOs
where recent collisions and fragmentation processes will have highly
modified their spins throughout the age of the solar system (Davis and
Farinella 1997).  The large volatile rich KBOs show significantly
different median period and possible amplitude rotational differences
when compared to the rocky large main belt asteroids which is expected
because of their differing compositions and collisional histories
(Sheppard and Jewitt 2002; Lacerda and Luu 2006).

I have furthered the photometric monitoring of large KBOs (absolute
magnitudes $H < 5.5$ or radii greater than about 100 km assuming
moderate albedos) in order to determine their short term rotational
and long term phase related light curves to better understand their
rotations, shapes and possible surface characteristics.  This is a
continuation of previous works (Jewitt and Sheppard 2002; Sheppard and
Jewitt 2002; Sheppard and Jewitt 2003; Sheppard and Jewitt 2004).

\section{Observations}

The data for this work were obtained at the Dupont 2.5 meter telescope
at Las Campanas in Chile and the University of Hawaii 2.2 meter
telescope atop Mauna Kea in Hawaii.

Observations at the Dupont 2.5 meter telescope were performed on the
nights of February 14, 15 and 16, March 9 and 10, October 25, 26, and
27, November 28, 29, and 30 and December 1, 2005 UT.  The instrument
used was the Tek5 with a $2048 \times 2048$ pixel CCD with $24$
$\micron$ pixels giving a scale of $0.\arcsec 259$ pixel$^{-1}$ at the
f/7.5 Cassegrain focus for a field of view of about $8\arcmin .85
\times 8\arcmin .85$.  Images were acquired through a Harris R-band
filter while the telescope was autoguided on nearby bright stars at
sidereal rates (Table 1).  Seeing was generally good and ranged from
$0.\arcsec 6$ to $1.\arcsec 5$ FWHM.

Observations at the University of Hawaii 2.2 meter telescope were
obtained on the nights of December 19, 21, 23 and 24, 2003 UT and used
the Tektronix $2048 \times 2048$ pixel CCD.  The pixels were $24$
$\micron$ in size giving $0.\arcsec 219$ pixel$^{-1}$ scale at the
f/10 Cassegrain focus for a field of view of about $7\arcmin .5 \times
7\arcmin .5$.  Images were obtained in the R-band filter based on the
Johnson-Kron-Cousins system with the telescope auto-guiding at
sidereal rates using nearby bright stars.  Seeing was very
good over the several nights ranging from $0.\arcsec 6$ to
$1.\arcsec 2$ FWHM.

For all observations the images were first bias subtracted and then
flat-fielded using the median of a set of dithered images of the
twilight sky.  The photometry for the KBOs was done in two ways in
order to optimize the signal-to-noise ratio.  First, aperture
correction photometry was performed by using a small aperture on the
KBOs ($0.\arcsec 65$ to $1.\arcsec 04$ in radius) and both the same
small aperture and a large aperture ($2.\arcsec 63$ to $3. \arcsec 63$
in radius) on several nearby unsaturated bright field stars.  The
magnitude within the small aperture used for the KBOs was corrected by
determining the correction from the small to the large aperture using
the PSF of the field stars.  Second, I performed photometry on the
KBOs using the same field stars but only using the large aperture on
the KBOs.  The smaller apertures allow better photometry for the
fainter objects since it uses only the high signal-to-noise central
pixels.  The range of radii varies because the actual radii used
depends on the seeing.  The worse the seeing the larger the radius of
the aperture needed in order to optimize the photometry.  Both
techniques found similar results, though as expected, the smaller
aperture gives less scatter for the fainter objects while the larger
aperture is superior for the brighter objects.

Photometric standard stars from Landolt (1992) were used for
calibration.  Each individual object was observed at all times in the
same filter and with the same telescope setup.  Relative photometric
calibration from night to night was very stable since the same fields
stars were observed.  The few observations that were taken in mildly
non-photometric conditions (i.e. thin cirrus) were easily calibrated
to observations of the same field stars on the photometric nights.
Thus, the data points on these mildly non-photometric nights are
almost as good as the other data with perhaps a slightly larger error
bar.  The dominate source of error in the photometry comes from
simple root N noise.

\section{Light Curve Causes}

The apparent magnitude or brightness of an atmospherless inert body in
our solar system is mainly from reflected sunlight and can be
calculated as
\begin{equation}
m_{R}=m_{\odot}-2.5\mbox{log}\left[p_{R}r^{2}\phi (\alpha )/(2.25\times 10^{16}R^{2}\Delta^{2})\right]  \label{eq:appmag}
\end{equation}
in which $r$ [km] is the radius of the KBO, $R$ [AU] is the
heliocentric distance, $\Delta$ [AU] is the geocentric distance,
$m_{\odot}$ is the apparent red magnitude of the sun ($-27.1$),
$m_{R}$ is the apparent red magnitude, $p_{R}$ is the red geometric
albedo, and $\phi (\alpha)$ is the phase function in which the phase
angle $\alpha=0$ deg at opposition and $\phi (0)=1$.

The apparent magnitude of the TNO may vary for the main following reasons:

1) The geometry in which $R, \Delta$ and/or $\alpha$ changes for the
TNO.  Geometrical considerations at the distances of the TNOs are
usually only noticeable over a few weeks or longer and thus are
considered long-term variations.  These are further discussed in
section 5.

2) The TNOs albedo, $p_{R}$, may not be uniform on its surface causing
   the apparent magnitude to vary as the different albedo markings on
   the TNOs surface rotate in and out of our line of sight.  Albedo or
   surface variations on an object usually cause less than a $30 \%$
   difference from maximum to minimum brightness of an object.
   (134340) Pluto, because of its atmosphere (Spencer et al. 1997),
   has one of the highest known amplitudes from albedo variations
   ($\sim 0.3$ magnitudes; Buie et al. 1997).

3) Shape variations or elongation of an object will cause the
   effective radius of an object to our line of sight to change as the
   TNO rotates.  A double peaked periodic light curve is expected to
   be seen in this case since the projected cross section would go
   between two minima (short axis) and two maxima (long axis) during
   one complete rotation of the TNO. Elongation from material strength
   is likely for small TNOs ($r < 100$ km) but for the larger TNOs
   observed in this paper no significant elongation is expected from
   material strength because their large self gravity.

   A large TNO ($r > 100$ km) may be significantly elongated if it has
   a large amount of rotational angular momentum.  An object will be
   near breakup if it has a rotation period near the critical rotation
   period ($P_{crit}$) at which centripetal acceleration equals
   gravitational acceleration towards the center of a rotating
   spherical object,
\begin{equation}
P_{crit} = \left(\frac{3\pi }{G \rho}\right)^{1/2}   \label{eq:equil}
\end{equation}
where $G$ is the gravitational constant and $\rho$ is the density of
the object.  With $\rho$ = $10^3$ kg m$^{-3}$ the critical period is
about 3.3 hours.  At periods just below the critical period the object
will likely break apart.  For objects with rotations significantly
above the critical period the shapes will be bimodal Maclaurin
spheroids which do not shown any significant rotational light curves
produced by shape (Jewitt and Sheppard 2002).  For periods just above
the critical period the equilibrium figures are triaxial ellipsoids
which are elongated from the large centripetal force and usually show
prominent rotational light curves (Weidenschilling 1981; Holsapple
2001; Jewitt and Sheppard 2002).

For an object that is triaxially elongated the peak-to-peak amplitude
of the rotational light curve allows for the determination of the
projection of the body shape into the plane of the sky by (Binzel et
al. 1989)
\begin{equation}
\Delta m=2.5\mbox{log}\left(\frac{a}{b}\right) - 1.25\mbox{log}\left(\frac{a^{2}cos^{2}\theta +c^{2}sin^{2}\theta}{b^{2}cos^{2}\theta +c^{2}sin
^{2}\theta}\right)   
\label{eq:elong}
\end{equation}
where $a \geq b \geq c$ are the semiaxes with the object in rotation
about the $c$ axis, $\Delta m$ is expressed in magnitudes, and
$\theta$ is the angle at which the rotation ($c$) axis is inclined to
the line of sight (an object with $\theta = 90$ deg. is viewed
equatorially).  The amplitudes of the light curves produced from
rotational elongation can range up to about 0.9 magnitudes (Leone et
al. 1984).

Assuming $\theta = 90$ degrees gives $a/b=10^{0.4 \Delta m}$.  Thus
the easily measured quantities of the rotation period and amplitude
can be used to determine a minimum density for an object if it is
assumed to be rotational elongated and strengthless (i.e. the bodies
structure behaves like a fluid, Chandrasekhar 1969).  The two best
cases of this high angular momentum elongation in the Kuiper Belt are
(20000) Varuna (Jewitt and Sheppard 2002) and (136108) 2003 EL$_{61}$
(Rabinowitz et al. 2006).

4) Periodic light curves may be produced if a TNO is an eclipsing or
   contact binary.  A double-peaked light curve would be expected with
   a possible characteristic notch shape near the minimum of the light
   curve.  Because the two objects may be tidally elongated the light
   curves can range up to about 1.2 magnitudes (Leone et al. 1984).
   The best example of such an object in the Kuiper Belt is 2001
   QG$_{298}$ (Sheppard and Jewitt 2004).

5) A non-periodic short-term light curve may occur from a complex
   rotational state, a recent collision, a binary with each component
   having a large light curve amplitude and a different rotation
   period or outgassing/cometary activity.  These types of short term
   variability are expected to be extremely rare and none have yet
   been reliably detected in the Kuiper Belt (Sheppard and Jewitt
   2003; Belskaya et al. 2006)

\section{Light Curve Results and Analysis}

The photometric measurements for the 10 newly observed KBOs are listed
in Table~1, where the columns include the start time of each
integration, the corresponding Julian date, and the magnitude.  No
correction for light travel time has been made.  Results of the
light curve analysis for all the KBOs newly observed are summarized in
Table~2.

The phase dispersion minimization (PDM) method (Stellingwerf 1978) was
used to search for periodicity in the individual light curves.  In
PDM, the metric is the so-called Theta parameter, which is essentially
the variance of the unphased data divided by the variance of the data
when phased by a given period.  The best fit period should have a very
small dispersion compared to the unphased data and thus Theta $<<$ 1
indicates that a good fit has been found.  In practice, a Theta less
than 0.4 indicates a possible periodic signature.

\subsection{(120348) 2004 TY$_{364}$}

\begin{figure}
\epsscale{0.7}
\centerline{\includegraphics[angle=90,width=\textwidth]{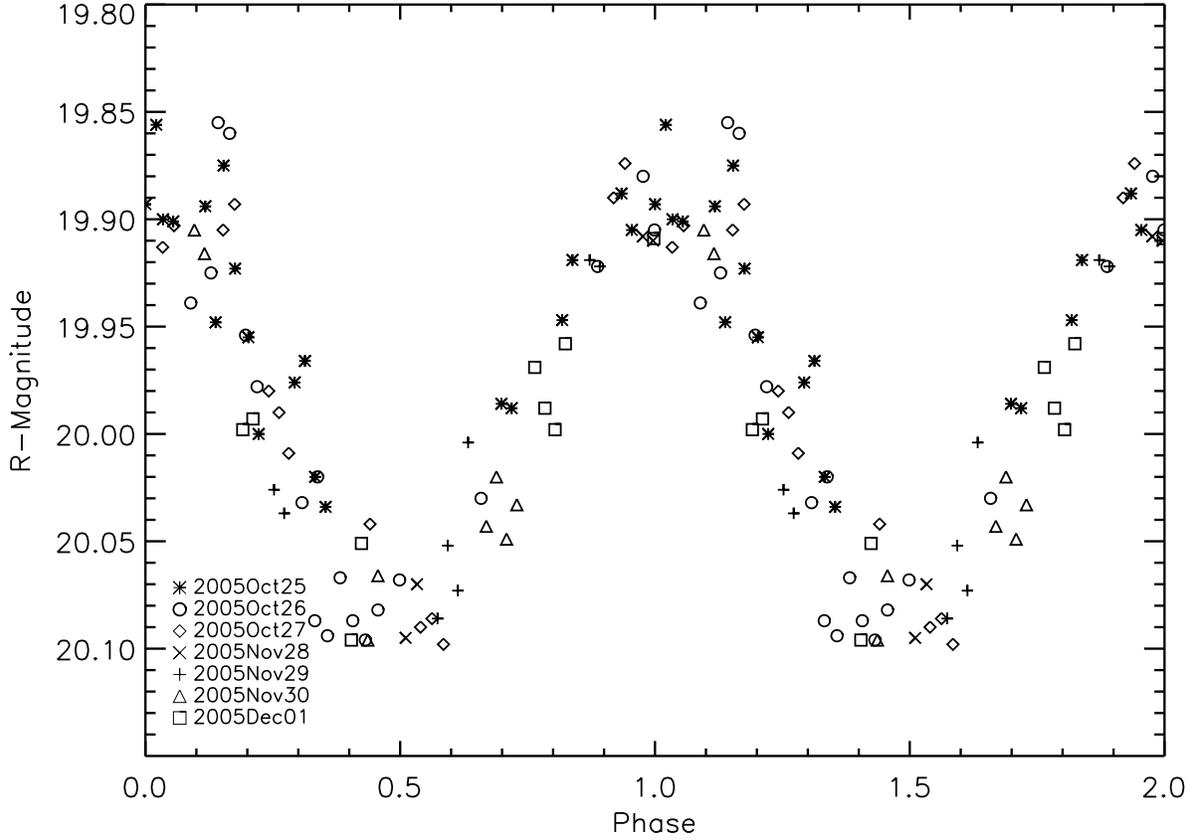}}
\caption{The phased best fit single-peaked period for (120348) 2004
  TY$_{364}$ of 5.85 hours.  The peak-to-peak amplitude is about 0.22
  magnitudes.  The data from November and December has been vertically
  shifted to correspond to the same phase angle as the data from
  October using the phase function found for this object in this work.
  Individual error bars for the measurements are not shown for clarity
  but are generally $\pm 0.01$ mags as seen in Table 1.}
\label{fig:phasesinglety} 
\end{figure}

\begin{figure}
\epsscale{0.7}
\centerline{\includegraphics[angle=90,width=\textwidth]{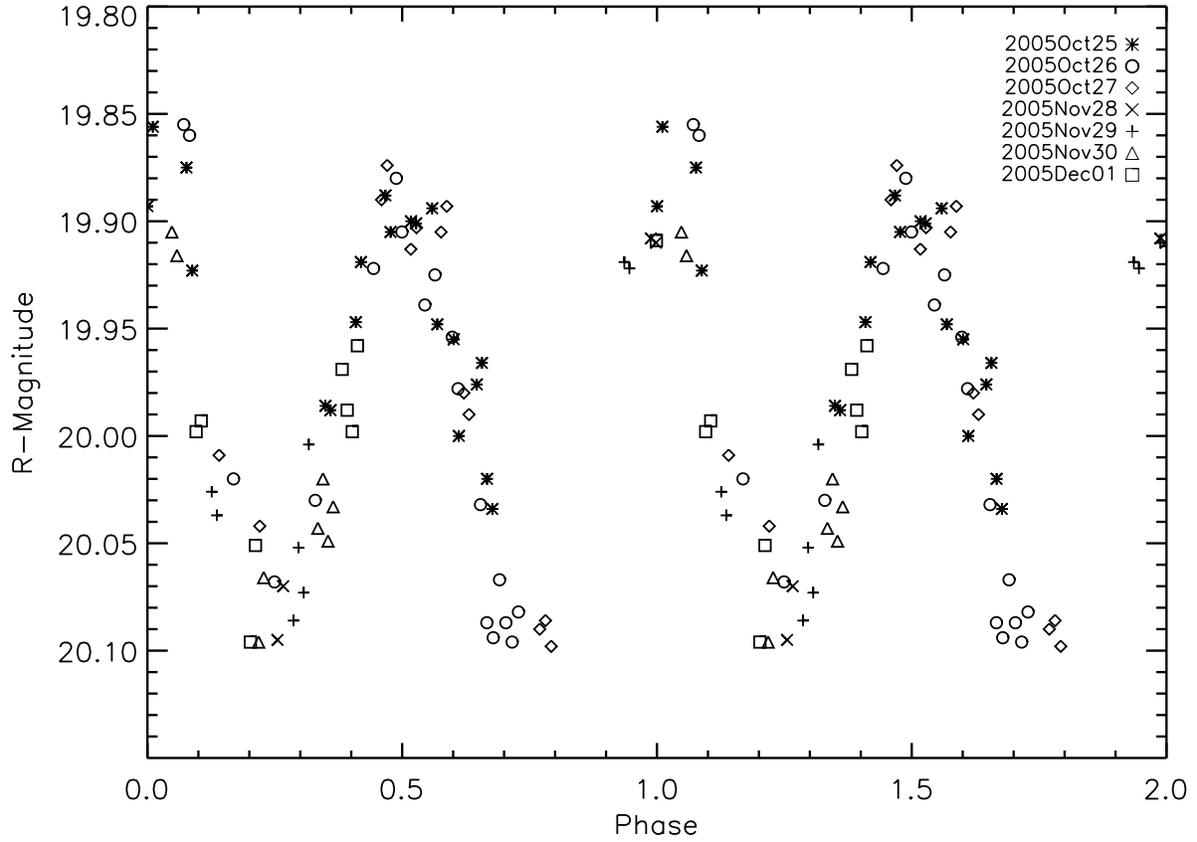}}
\caption{The phased double-peaked period for (120348) 2004 TY$_{364}$
  of 11.70 hours.  The data from November and December has been
  vertically shifted to correspond to the same phase angle as the data
  from October using the phase function found for this object in this
  work.  Individual error bars for the measurements are not shown for
  clarity but are generally $\pm 0.01$ mags as seen in Table 1.}
\label{fig:phasedoublety} 
\end{figure}

Through the PDM analysis I found a strong Theta minima for 2004
TY$_{364}$ near a period of $P=5.85$ hours with weaker alias periods
flanking this (Figure~\ref{fig:pdmty}).  Phasing the data to all
possible periods in the PDM plot with Theta $< 0.4$ found that only
the single-peaked period near 5.85 hours and the double-peaked period
near 11.70 hours fits all the data obtained from October, November and
December 2005.  Both periods have an equally low Theta parameter of
about 0.15 and either could be the true rotation period
(Figures~\ref{fig:phasesinglety} and~\ref{fig:phasedoublety}).  The
peak-to-peak amplitude is $0.22\pm 0.02$ magnitudes.

If 2004 TY$_{364}$ has a double-peaked period it may be elongated from
its high angular momentum.  If the TNO is assumed to be observed
equator on then from Equation~\ref{eq:elong} the $a:b$ axis ratio is
about 1.2.  Following Jewitt and Sheppard (2002) I assume the TNO is a
rotationally elongated strengthless rubble pile.  Using the spin
period of 11.7 hours, the 1.2 $a:b$ axis ratio found above and the
Jacobi ellipsoid tables produced by Chandrasekhar (1969) I find the
minimum density of 2004 TY$_{364}$ is about 290 kg m$^{-3}$ with an
$a:c$ axis ratio of about 1.9.  This density is quite low which leads
one to believe either the TNO is not being viewed equator on or the
relatively long double-peaked period is not created from high angular
momentum of the object.

\subsection{(84922) 2003 VS$_{2}$}

\begin{figure}
\epsscale{0.7}
\centerline{\includegraphics[angle=90,width=\textwidth]{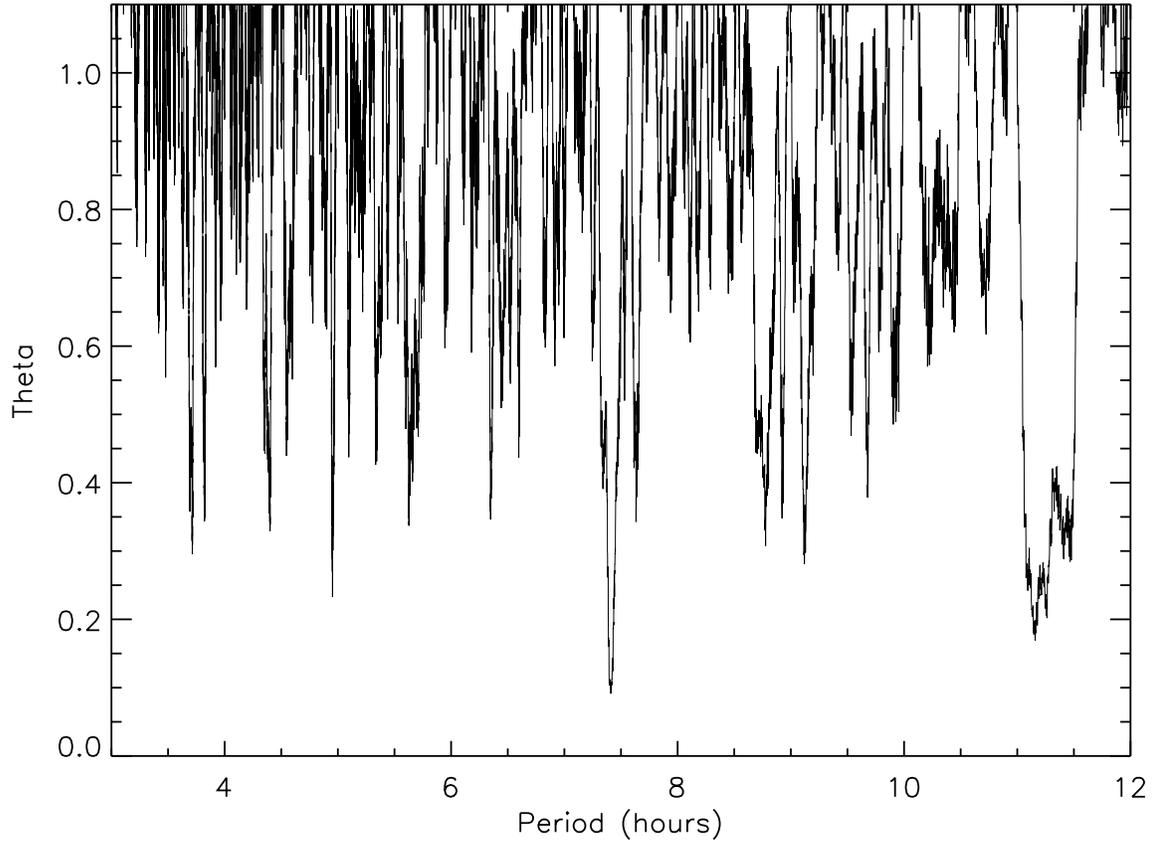}}
\caption{The Phase Dispersion Minimization (PDM) plot for (84922) 2003
VS$_{2}$.  The best fit is the double-peaked period near 7.41 hours.}
\label{fig:pdmvs} 
\end{figure}

\begin{figure}
\epsscale{0.7}
\centerline{\includegraphics[angle=90,width=\textwidth]{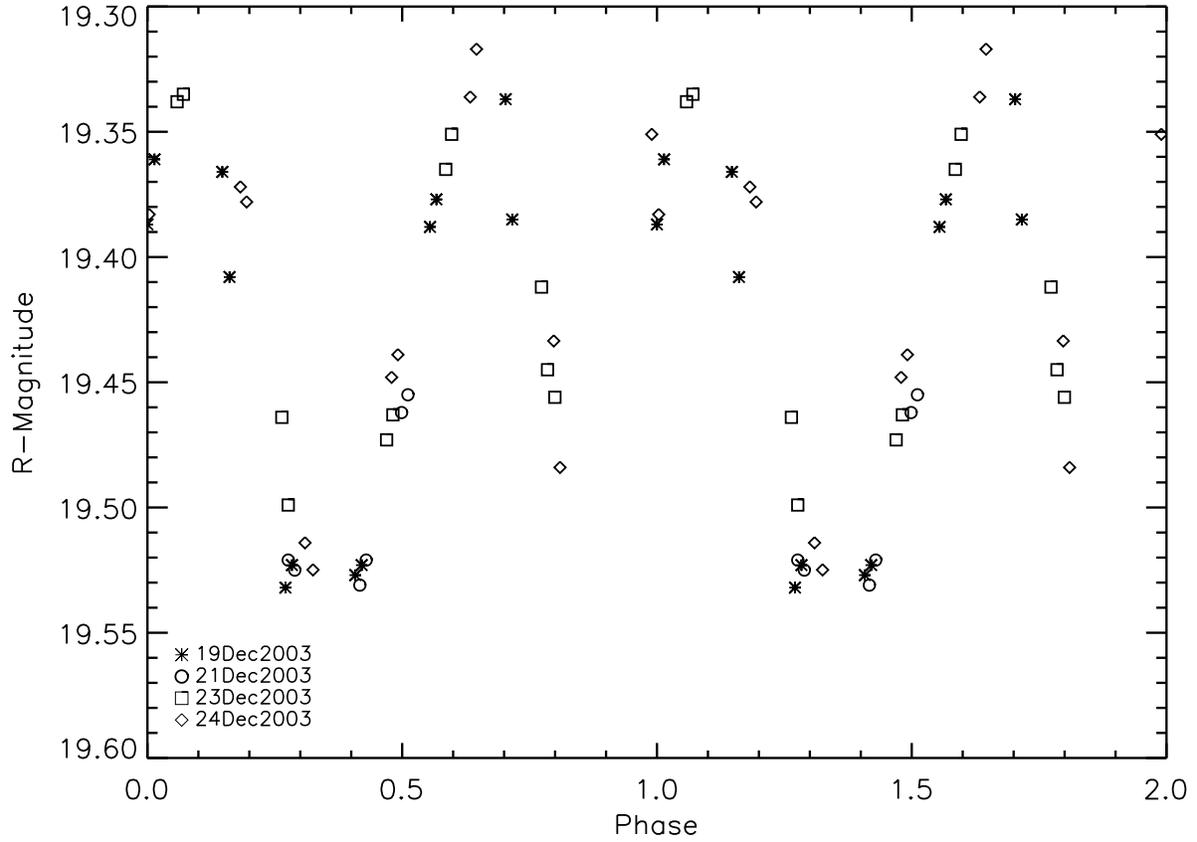}}
\caption{The phased best fit double-peaked period for (84922) 2003
  VS$_{2}$ of 7.41 hours.  The peak-to-peak amplitude is about 0.21
  magnitudes.  The two peaks have differences since one is slightly
  wider while the other is slightly shorter in amplitude.  This is the
  best fit period for (84922) 2003 VS$_{2}$.  Individual error bars
  for the measurements are not shown for clarity but are generally
  $\pm 0.01$ mags as seen in Table 1.}
\label{fig:phasedoublevs} 
\end{figure}

\begin{figure}
\epsscale{0.7}
\centerline{\includegraphics[angle=90,width=\textwidth]{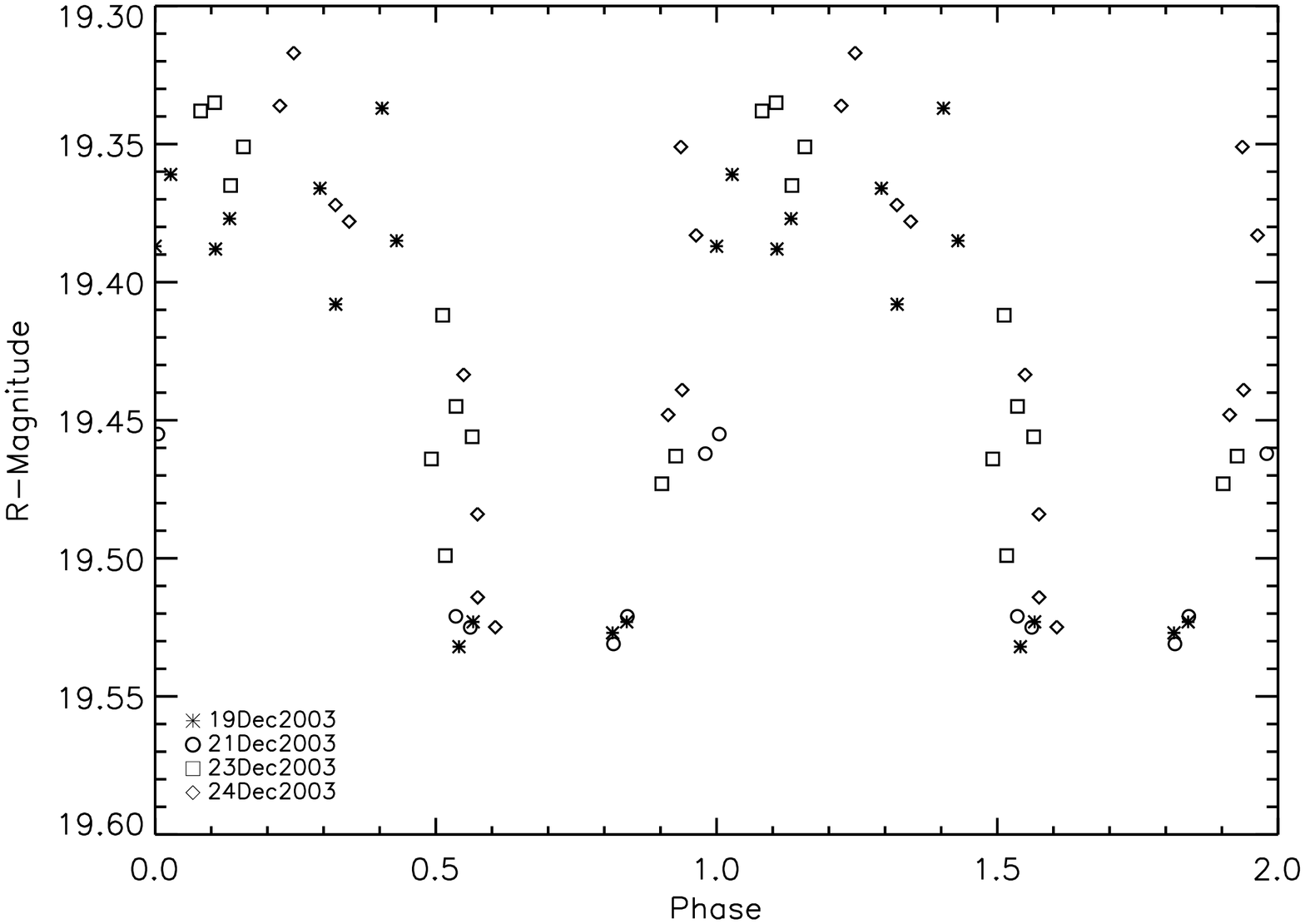}}
\caption{The phased single-peaked period for (84922) 2003 VS$_{2}$ of
  3.70 hours.  The single peaked period for 2003 VS$_{2}$ does not
  look well matched and has a larger scatter about the solution
  compared to the double-peaked period shown in
  Figure~\ref{fig:phasedoublevs}.  Individual error bars for the
  measurements are not shown for clarity but are generally $\pm 0.01$
  mags as seen in Table 1.}
\label{fig:phasesinglevs} 
\end{figure}

\begin{figure}
\epsscale{0.7}
\centerline{\includegraphics[angle=90,width=\textwidth]{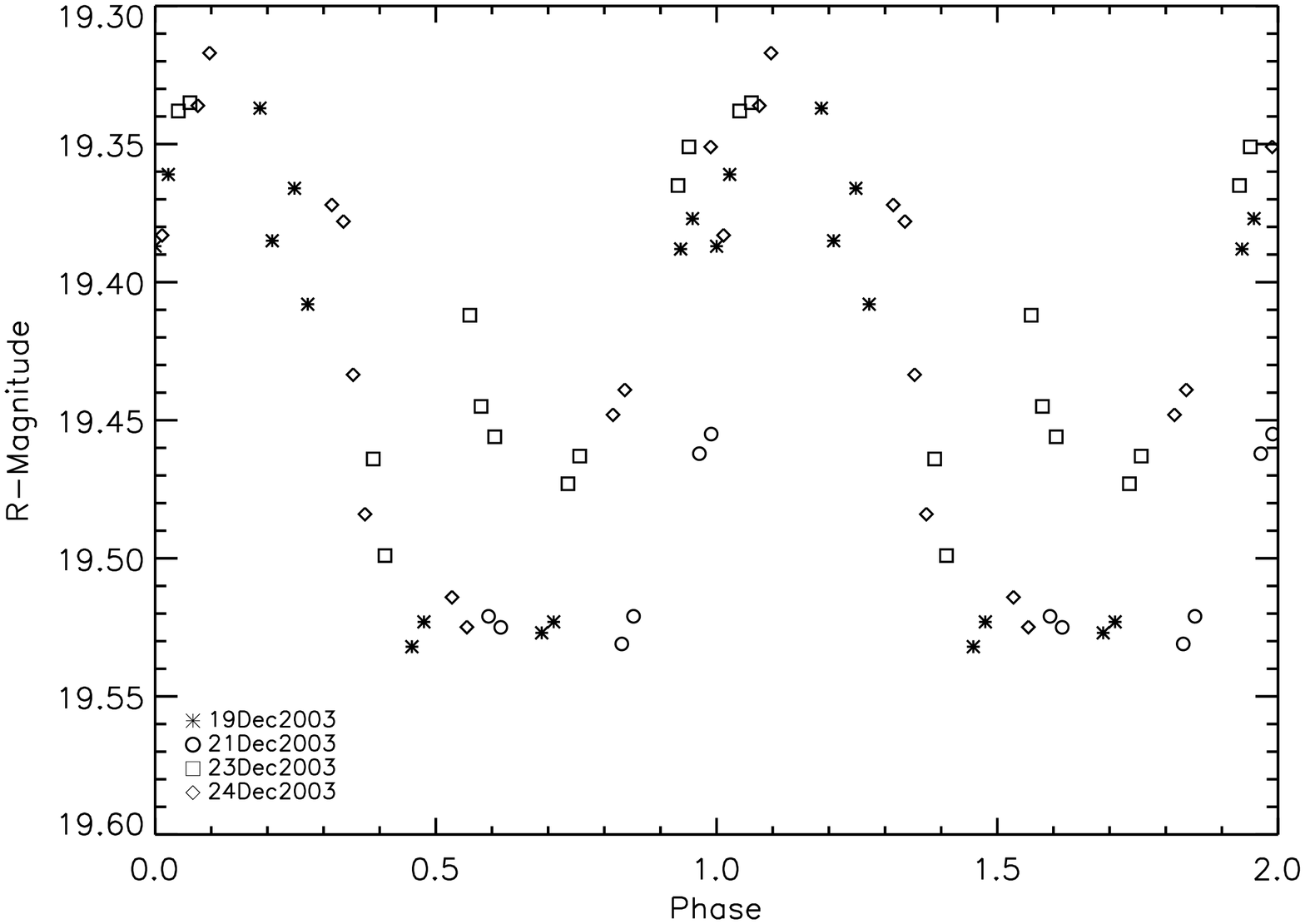}}
\caption{The phased single-peaked period for (84922) 2003 VS$_{2}$ of
  4.39 hours.  Again, the single peaked period for 2003 VS$_{2}$ does
  not look well matched and has a larger scatter about the solution
  compared to the double-peaked period shown in
  Figure~\ref{fig:phasedoublevs}.  Individual error bars for the
  measurements are not shown for clarity but are generally $\pm 0.01$
  mags as seen in Table 1.}
\label{fig:phasesinglevs2} 
\end{figure}

\begin{figure}
\epsscale{0.7}
\centerline{\includegraphics[angle=90,width=\textwidth]{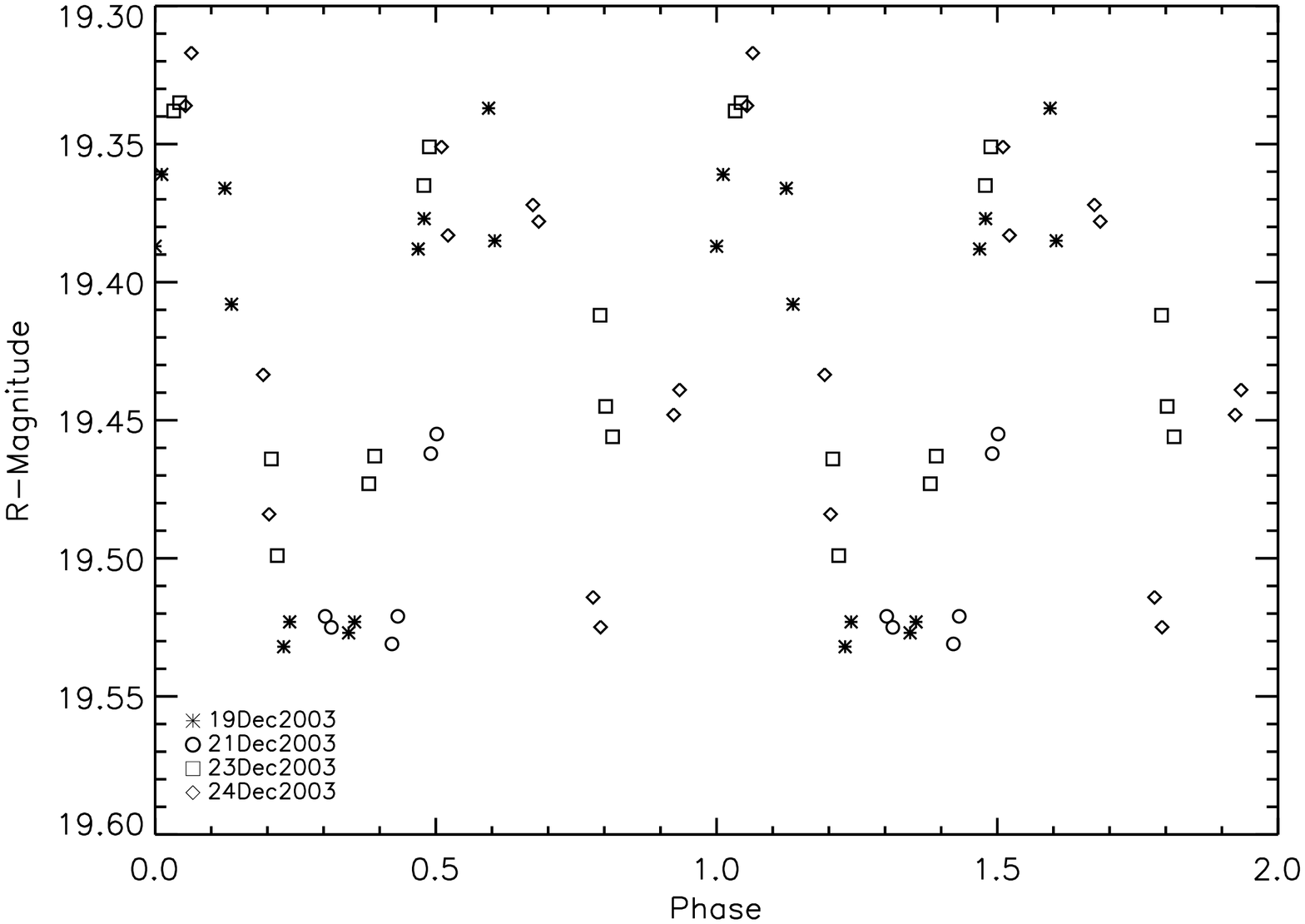}}
\caption{The phased double-peaked period for (84922) 2003 VS$_{2}$ of
  8.77 hours.  This double-peaked period for 2003 VS$_{2}$ does not
  look well matched and has a larger scatter about the solution
  compared to the 7.41 hour double-peaked period shown in
  Figure~\ref{fig:phasedoublevs}.  Individual error bars for the
  measurements are not shown for clarity but are generally $\pm 0.01$
  mags as seen in Table 1.}
\label{fig:phasedoublevs2} 
\end{figure}

The KBO 2003 VS$_{2}$ has a very low Theta of less than 0.1 near 7.41
hours in the PDM plot (Figure~\ref{fig:pdmvs}).  Phasing the December
2003 data to this period shows a well defined double-peaked period
(Figure~\ref{fig:phasedoublevs}).  The single peaked period for this
result would be near 3.71 hours which was a possible period determined
for this object by Ortiz et al. (2006).  The 3.71 hour single-peaked
period does not look as convincing (Figure~\ref{fig:phasesinglevs})
which confirms the PDM result that the single-peaked period has about
three times more dispersion than the double-peaked period.  This is
likely because one of the peaks is taller in amplitude ($\sim 0.05$
mags) and a little wider.  The other single-peaked period of 4.39
hours (Figure~\ref{fig:phasesinglevs2}) and the double-peaked period
of 8.77 hours (Figure~\ref{fig:phasedoublevs2}) mentioned by Oritz et
al. (2006) do not show a low Theta in the PDM and also do not look
convincing when examining the phased data.  The peak-to-peak
amplitude is $0.21\pm 0.02$ magnitudes, which is similar to that
detected by Ortiz et al. (2006).

The fast rotation of 7.41 hours and double-peaked nature suggests that
2003 VS$_{2}$ may be elongated from its high angular momentum.  Using
Equation~\ref{eq:elong} and assuming the TNO is observed equator on
the $a:b$ axis ratio is about 1.2.  Using the spin period of 7.41
hours, the 1.2 $a:b$ axis ratio and the Jacobi ellipsoid tables
produced by Chandrasekhar (1969) I find the minimum density of 2003
VS$_{2}$ is about 720 kg m$^{-3}$ with an $a:c$ axis ratio of about
1.9.  This result is similar to other TNO densities found through the
Jacobian Ellipsoid assumption (Jewitt and Sheppard 2002; Sheppard and
Jewitt 2002; Rabinowitz et al. 2006) as well as recent thermal results
from the Spitzer space telescope (Stansberry et al. 2006).

\subsection{(126154) 2001 YH$_{140}$}

\begin{figure}
\epsscale{0.7}
\centerline{\includegraphics[angle=90,width=\textwidth]{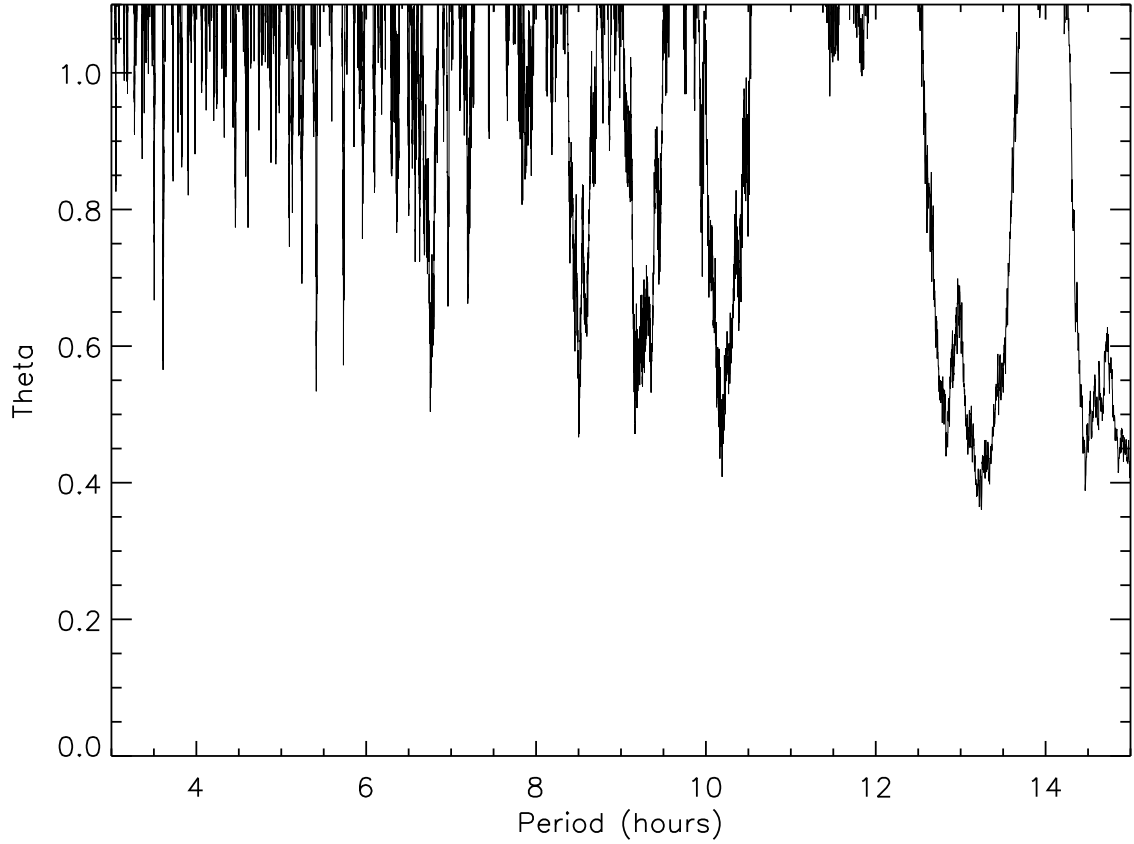}}
\caption{The Phase Dispersion Minimization (PDM) plot for 2001 YH$_{140}$.
The best fit is the single-peaked period near 13.25 hours.  The other
possible fits near 8.5, 9.15 and 10.25 hours don't look good when
phasing the data and viewing the result by eye.}
\label{fig:pdmyh} 
\end{figure}

\begin{figure}
\epsscale{0.7}
\centerline{\includegraphics[angle=90,width=\textwidth]{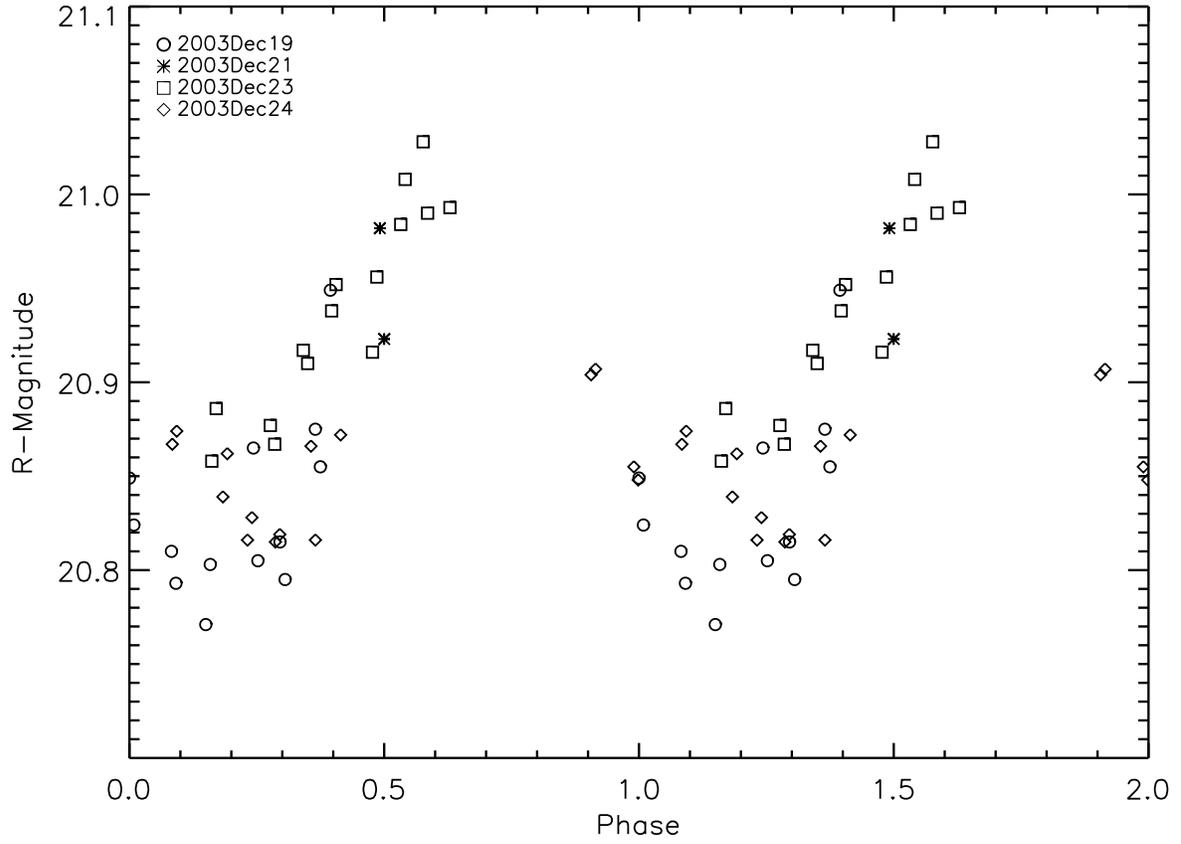}}
\caption{The phased best fit single-peaked period for 2001 YH$_{140}$
  of 13.25 hours.  The peak-to-peak amplitude is about 0.21
  magnitudes.  Individual error bars for the measurements are not
  shown for clarity but are generally $\pm 0.02$ mags as seen in Table
  1.}
\label{fig:phasesingleyh} 
\end{figure}

(126154) 2001 YH$_{140}$ shows variability of $0.21 \pm 0.04$
magnitudes.  The PDM for this TNO shows possible periods near 8.5,
9.15, 10.25 and 13.25 hours though only the 13.25 hour period has a
Theta less than 0.4 (Figure~\ref{fig:pdmyh}).  Visibly examining the
phased data finds only the 13.25 hour period is viable
(Figure~\ref{fig:phasesingleyh}).  This is consistent with the
observation that one minimum and one maximum were shown on December
23, 2003 in about six and a half hours, which would give a
single-peaked light curve of twice this time or about 13.25 hours.
Ortiz et al. (2006) found this object to have a similar variability
but with very limited data could not obtain a reliable period.  Ortiz
et al. did have one period of 12.99 hours which may be consistent with
our result.

\subsection{Flat Rotation Curves}

\begin{figure}
\epsscale{0.7}
\centerline{\includegraphics[angle=90,width=\textwidth]{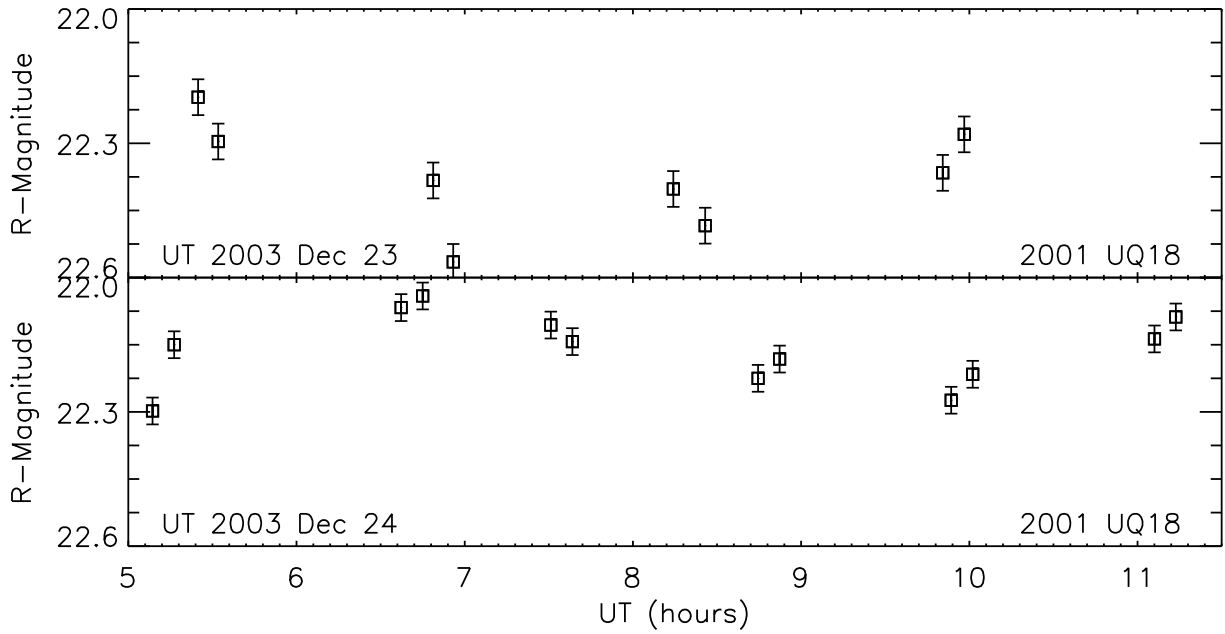}}
\caption{The flat light curve of 2001 UQ$_{18}$.  The KBO may have a
significant amplitude light curve but further observations are needed
to confirm.}
\label{fig:multiuq} 
\end{figure}

\begin{figure}
\epsscale{0.7}
\centerline{\includegraphics[angle=90,width=\textwidth]{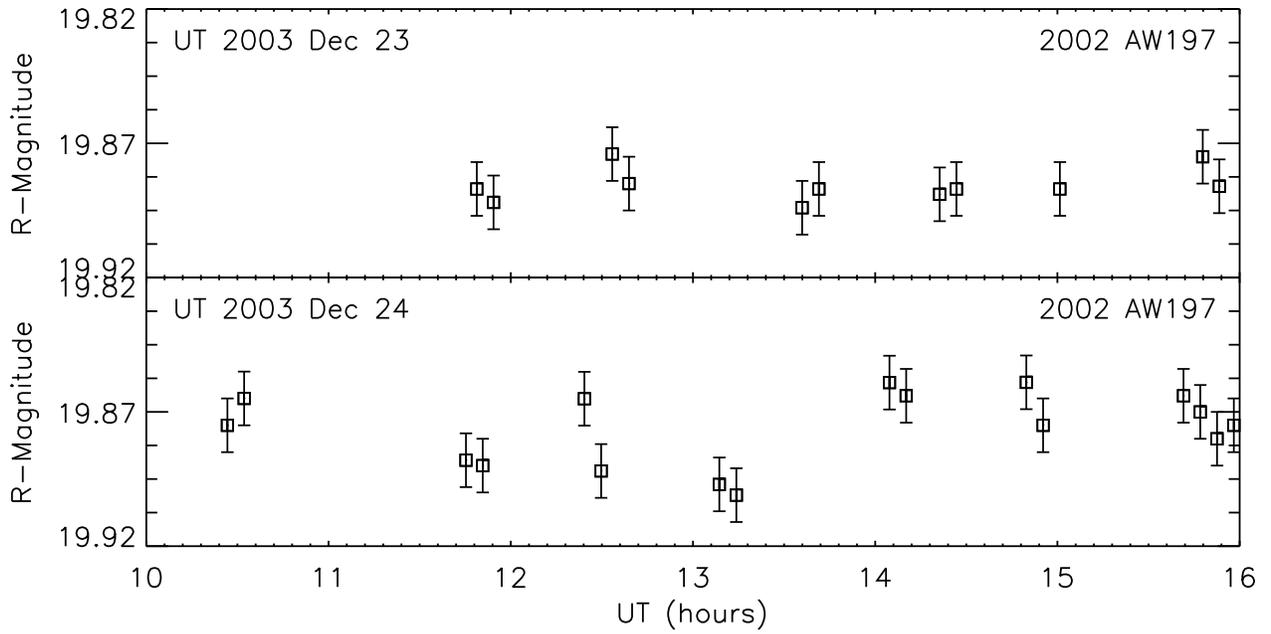}}
\caption{The flat light curve of (55565) 2002 AW$_{197}$.  The KBO has no
significant short-term variations larger than 0.03 magnitudes over two
days.}
\label{fig:multiaw} 
\end{figure}

\begin{figure}
\epsscale{0.7}
\centerline{\includegraphics[angle=90,width=\textwidth]{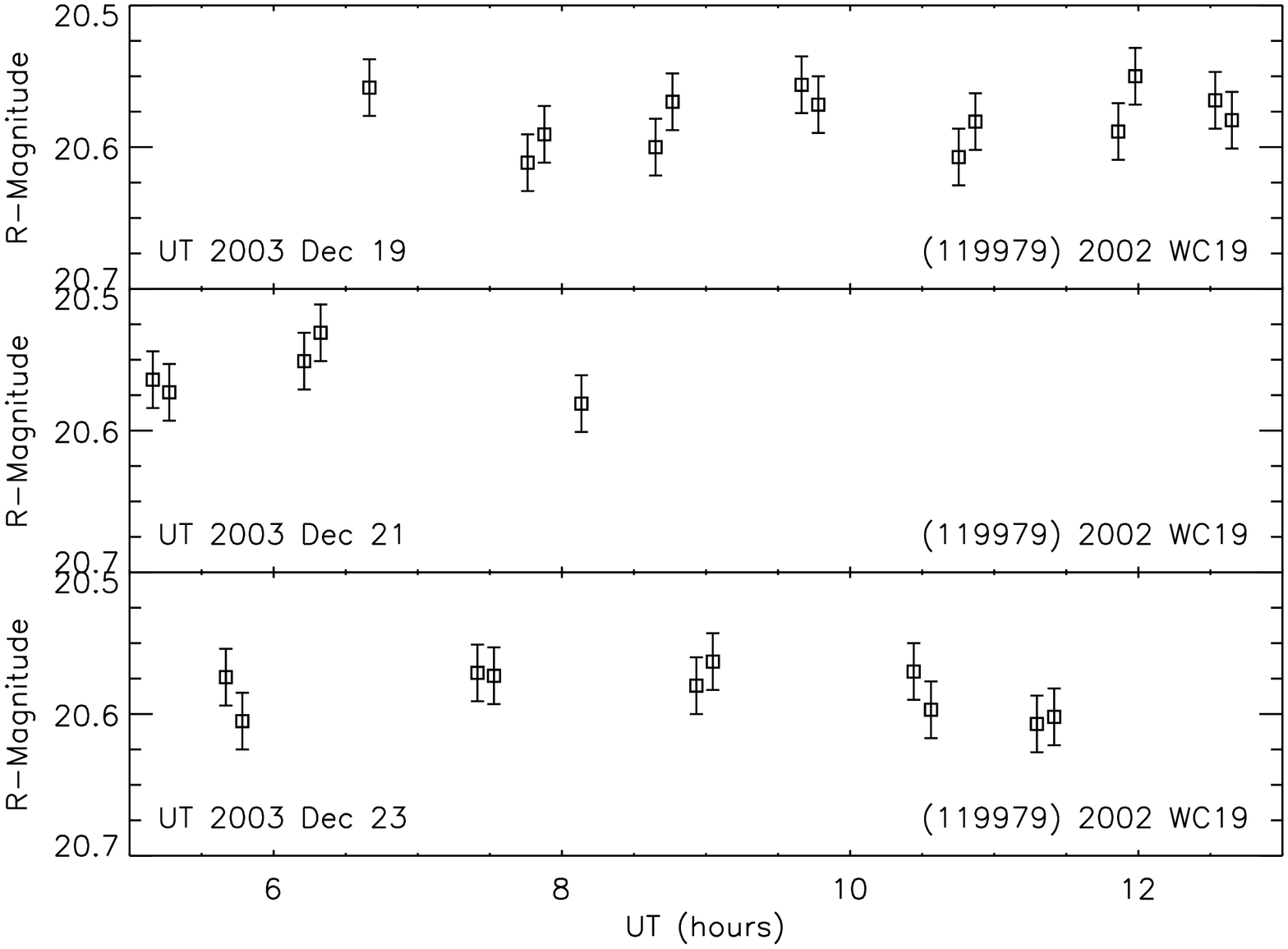}}
\caption{The flat light curve of (119979) 2002 WC$_{19}$.  The KBO has no
significant short-term variations larger than 0.03 magnitudes over four
days.}
\label{fig:multiwc} 
\end{figure}

\begin{figure}
\epsscale{0.7}
\centerline{\includegraphics[angle=90,width=\textwidth]{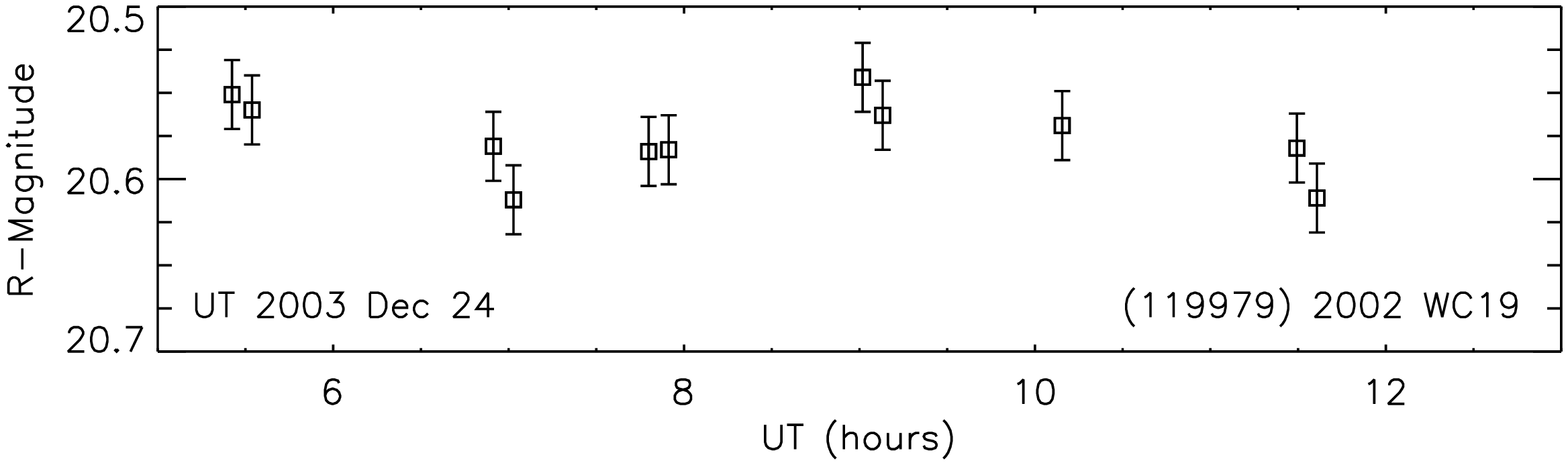}}
\caption{The flat light curve of (119979) 2002 WC$_{19}$.  The KBO has no
significant short-term variations larger than 0.03 magnitudes over four
days.}
\label{fig:multiwcb} 
\end{figure}

\begin{figure}
\epsscale{0.7}
\centerline{\includegraphics[angle=90,width=\textwidth]{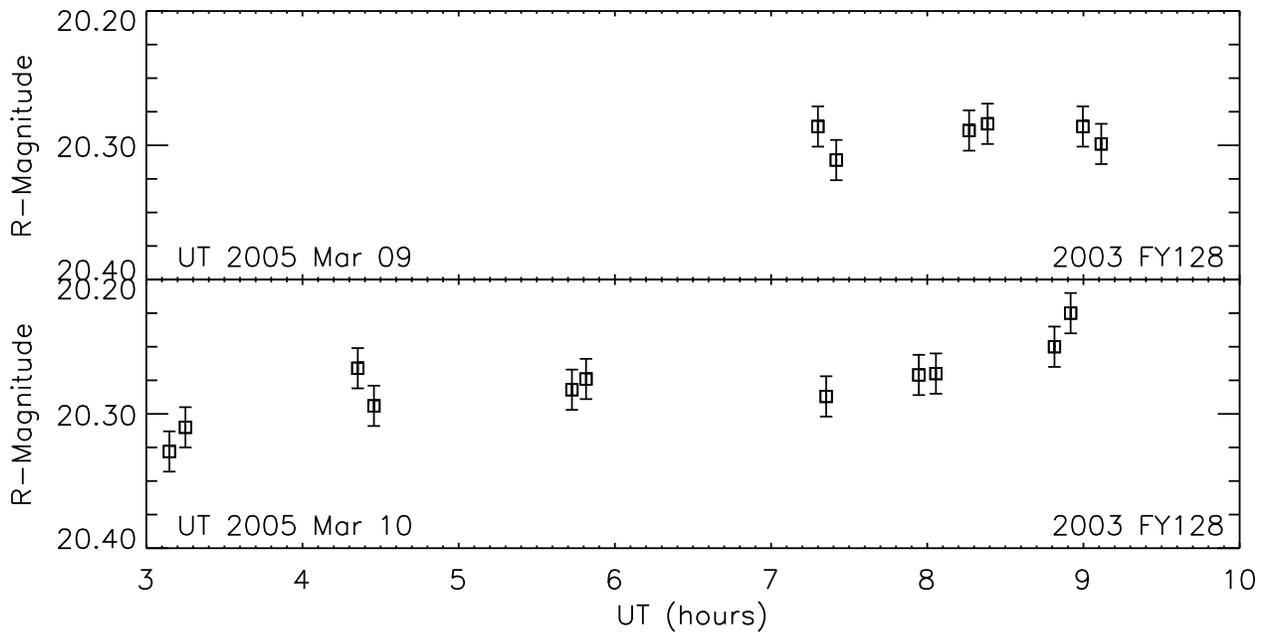}}
\caption{The flat light curve of (120132) 2003 FY$_{128}$.  The KBO has no
significant short-term variations larger than 0.08 magnitudes over two
days.}
\label{fig:multify} 
\end{figure}

\begin{figure}
\epsscale{0.7}
\centerline{\includegraphics[angle=90,width=\textwidth]{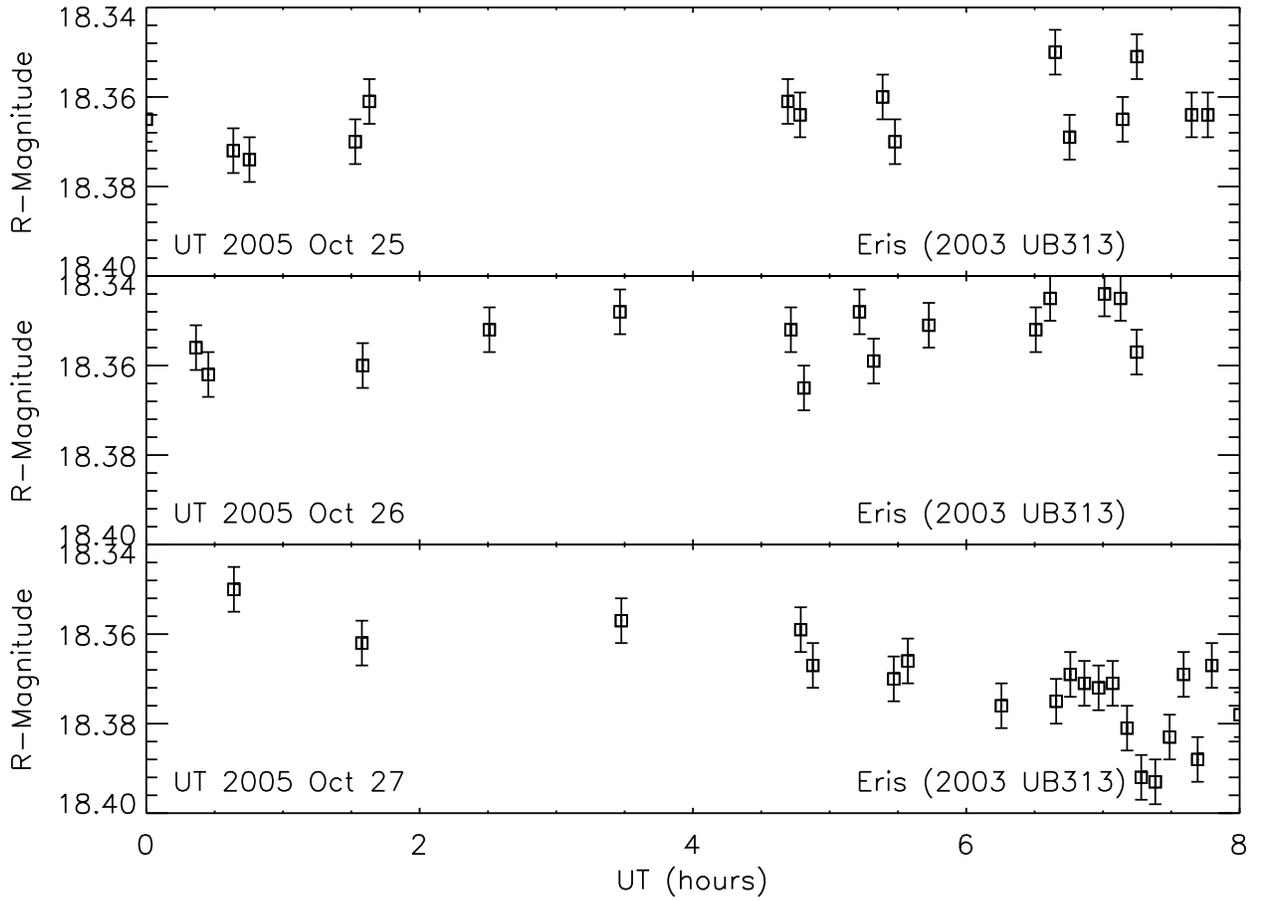}}
\caption{The flat light curve of Eris (2003 UB$_{313}$) in October 2005.  The KBO has no
significant short-term variations larger than 0.01 magnitudes over
several days.}
\label{fig:multiub1} 
\end{figure}

\begin{figure}
\epsscale{0.7}
\centerline{\includegraphics[angle=90,width=\textwidth]{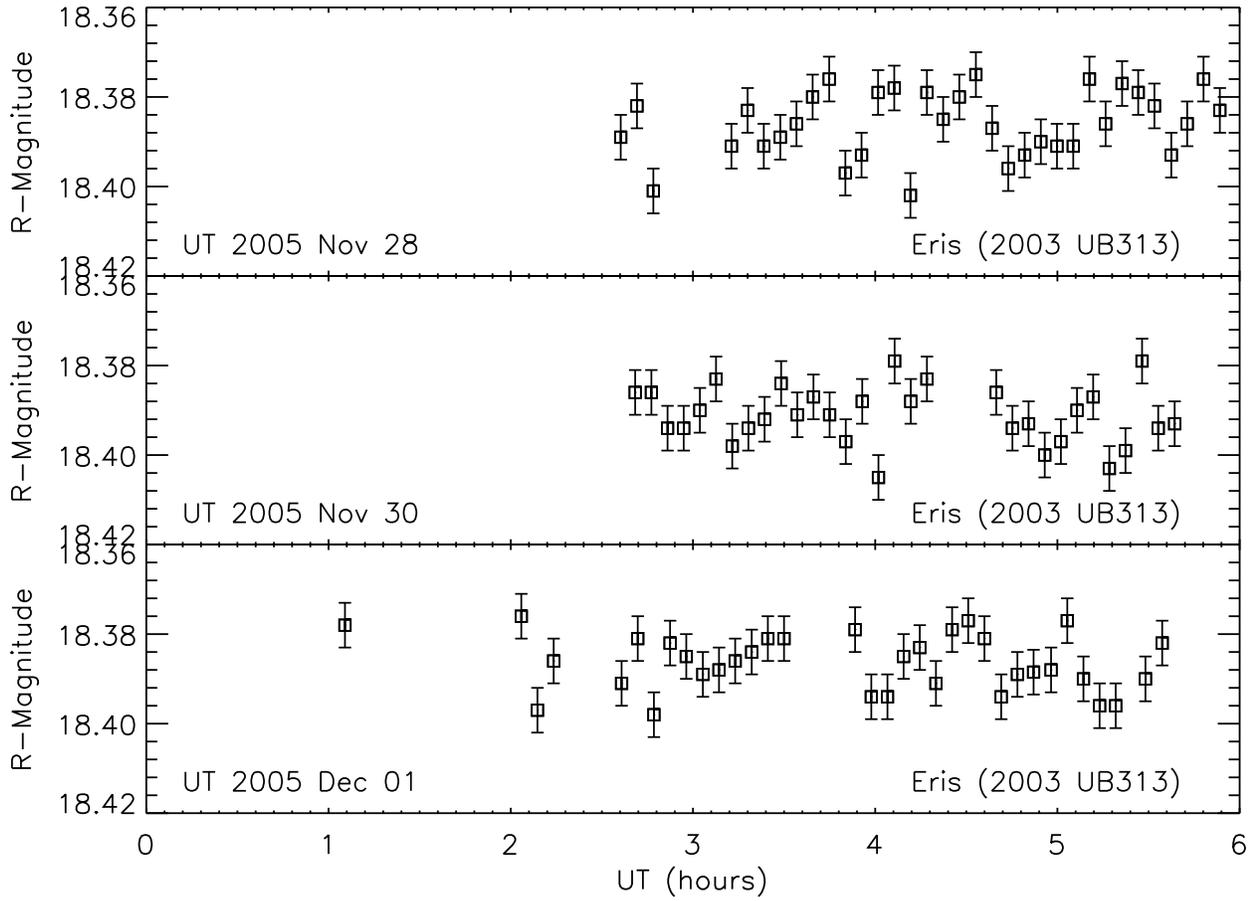}}
\caption{The flat light curve of Eris (2003 UB$_{313}$) in November and December
2005.  The KBO has no significant short-term variations larger than
0.01 magnitudes over several days.}
\label{fig:multiub2} 
\end{figure}

\begin{figure}
\epsscale{0.7}
\centerline{\includegraphics[angle=90,width=\textwidth]{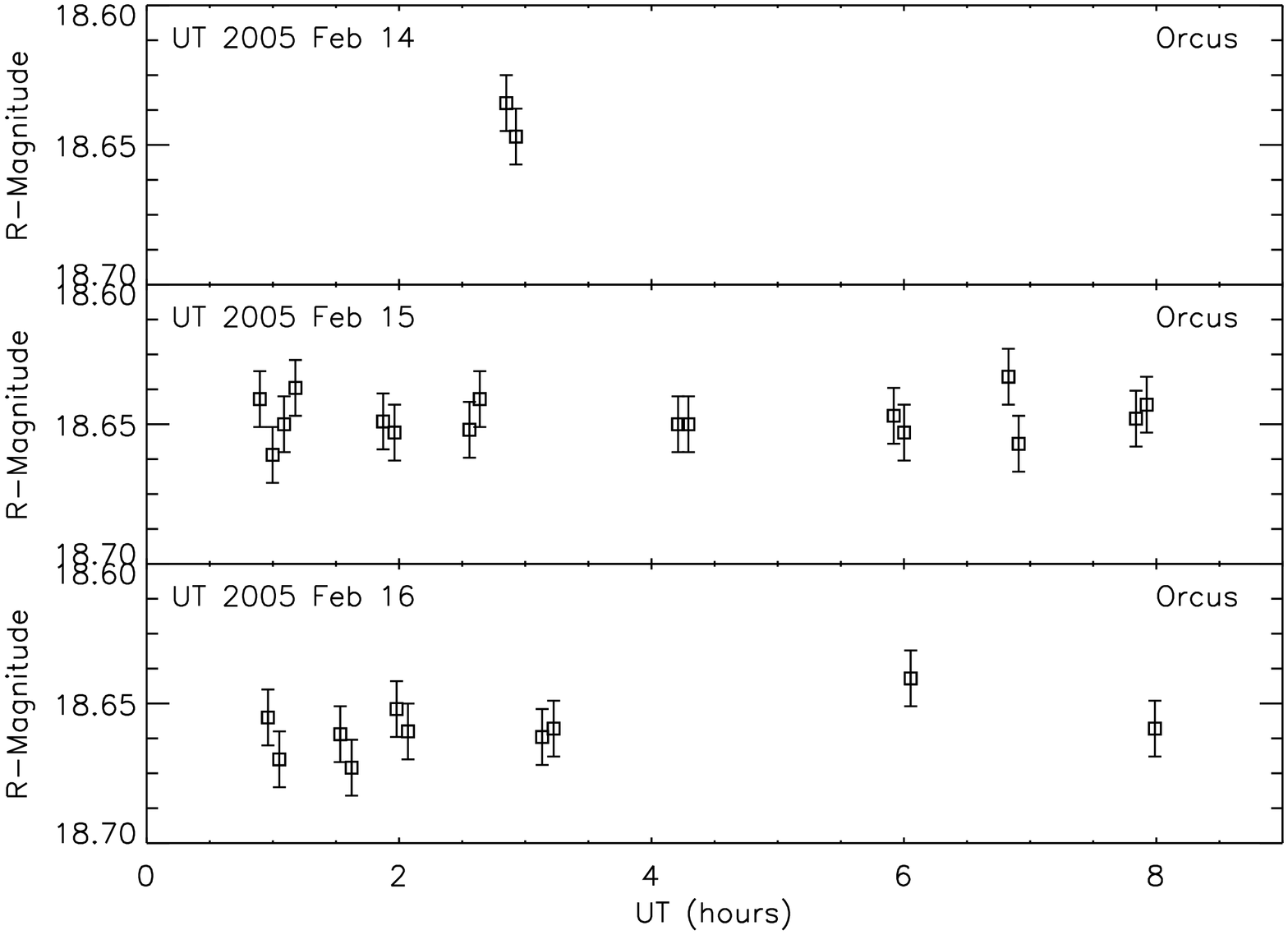}}
\caption{The flat light curve of (90482) Orcus 2004 DW in February
  2005.  The KBO has no significant short-term variations larger than
  0.03 magnitudes over several days.}
\label{fig:multidw1} 
\end{figure}

\begin{figure}
\epsscale{0.7}
\centerline{\includegraphics[angle=90,width=\textwidth]{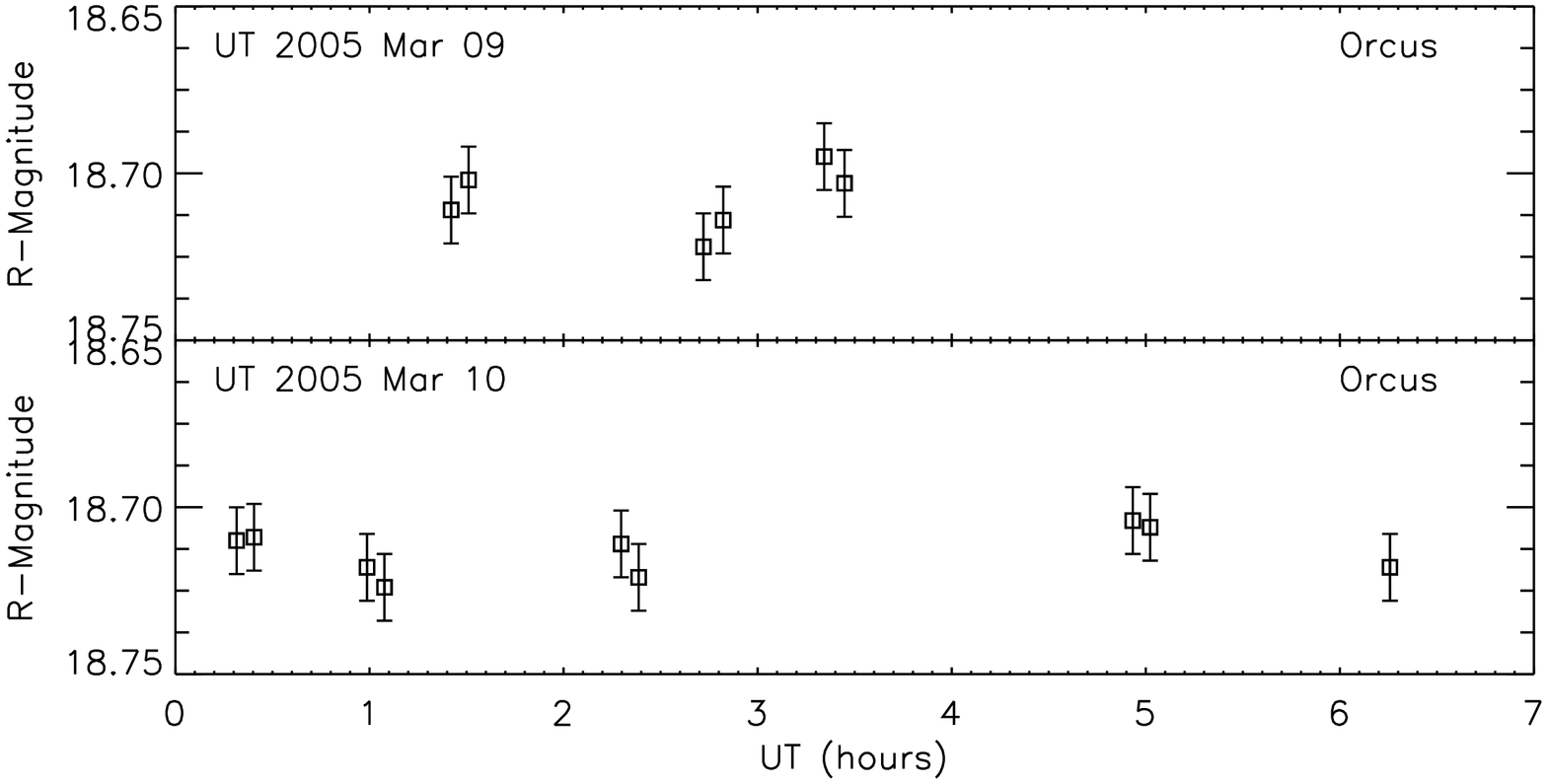}}
\caption{The flat light curve of (90482) Orcus 2004 DW in March 2005.
  The KBO has no significant short-term variations larger than 0.03
  magnitudes over several days.}
\label{fig:multidw2} 
\end{figure}

\begin{figure}
\epsscale{0.7}
\centerline{\includegraphics[angle=90,width=\textwidth]{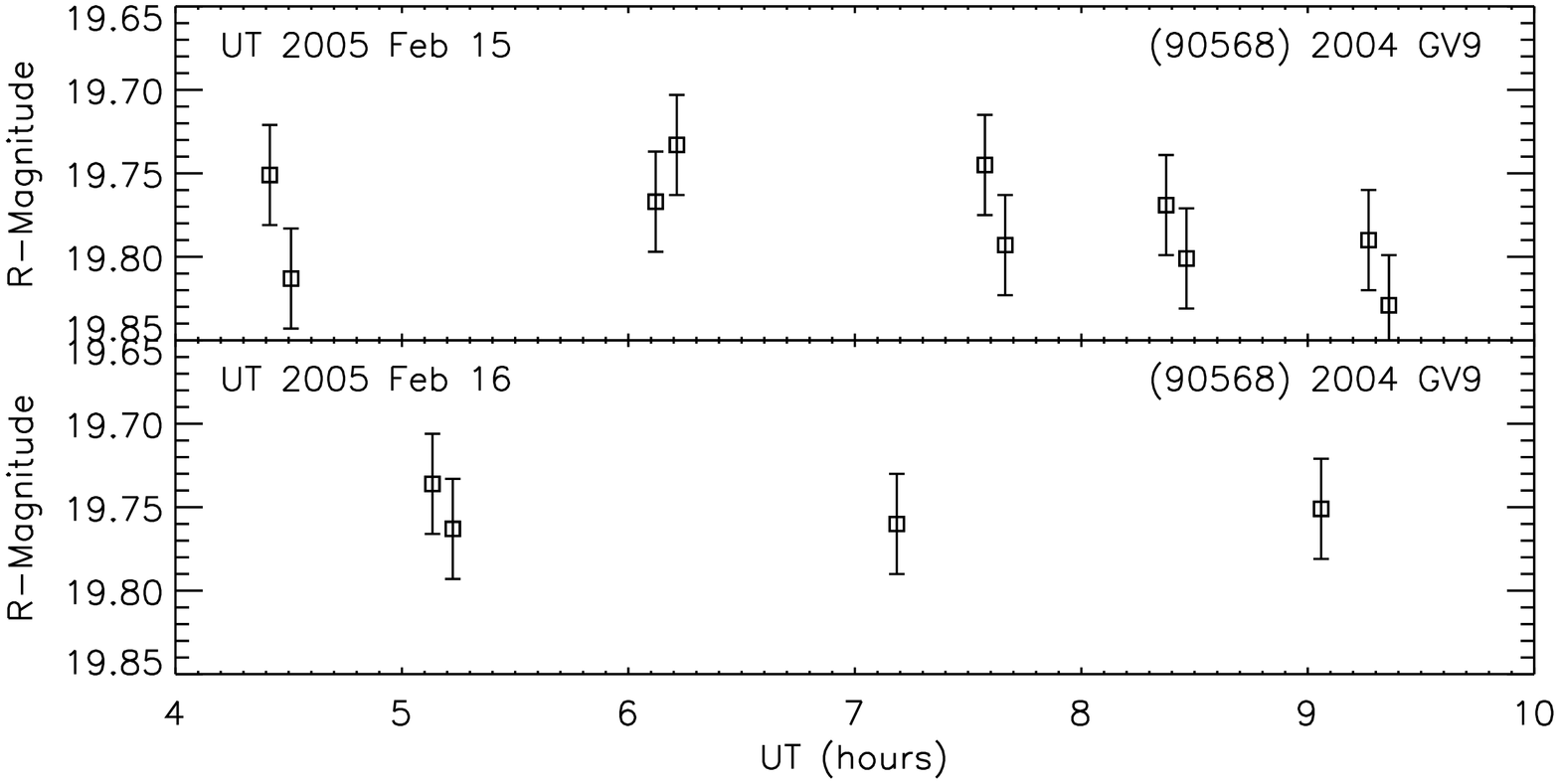}}
\caption{The flat light curve of (90568) 2004 GV$_{9}$ in February
  2005.  The KBO has no significant short-term variations larger than
  0.1 magnitudes over several days.}
\label{fig:multigv1} 
\end{figure}

\begin{figure}
\epsscale{0.7}
\centerline{\includegraphics[angle=90,width=\textwidth]{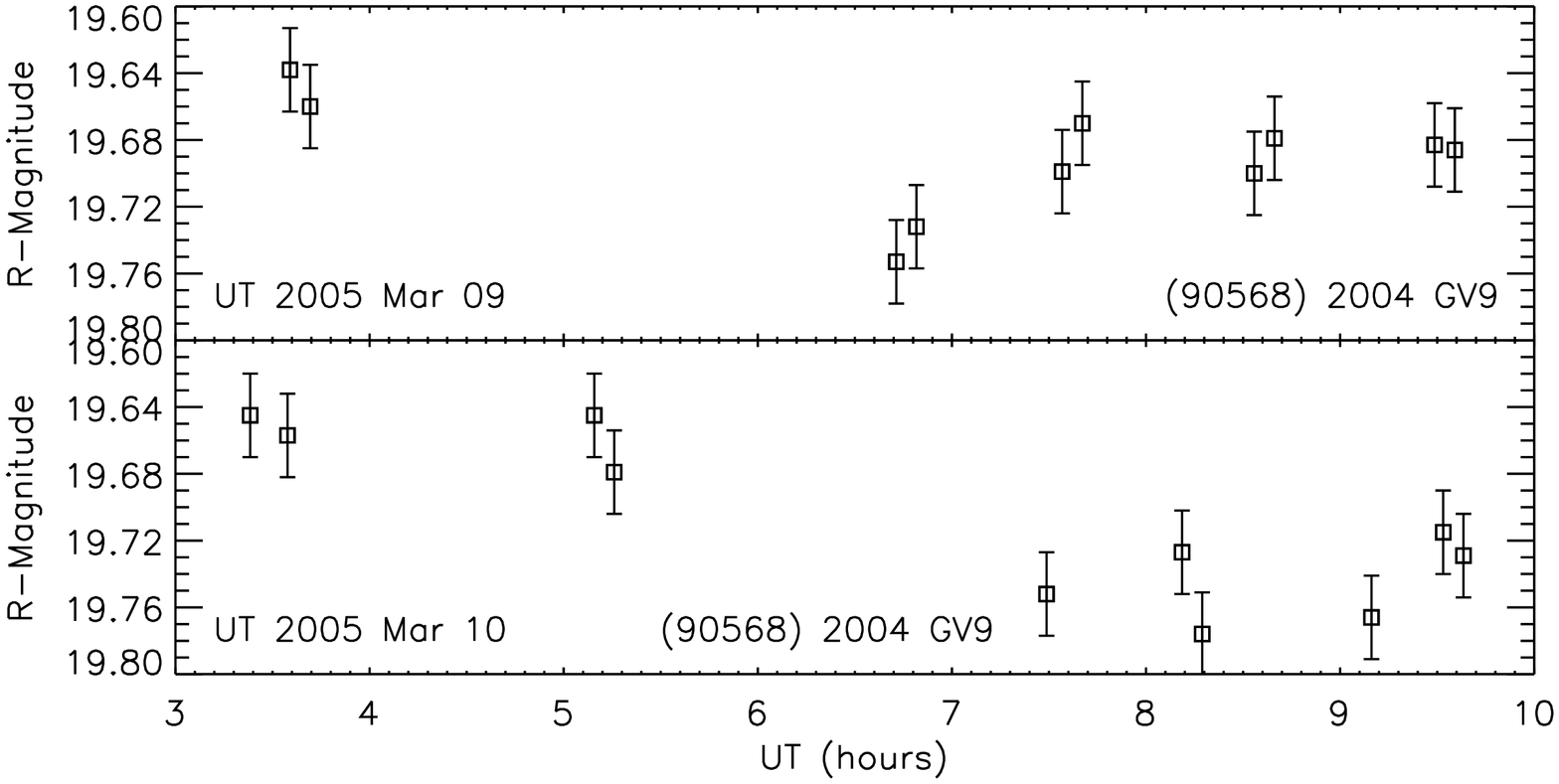}}
\caption{The flat light curve of (90568) 2004 GV$_{9}$ in March 2005.
  The KBO has no significant short-term variations larger than 0.1
  magnitudes over several days.}
\label{fig:multigv2} 
\end{figure}

Seven of the ten newly observed KBOs; 2001 UQ$_{18}$, (55565) 2002
AW$_{197}$, (119979) 2002 WC$_{19}$, (120132) 2003 FY$_{128}$,
(136199) Eris 2003 UB$_{313}$, (90482) Orcus 2004 DW, and (90568) 2004
GV$_{9}$ showed no variability within the photometric uncertainties of
the observations (Table 2; Figures~\ref{fig:multiuq}
to~\ref{fig:multigv2}).  These KBOs thus either have extremely long
rotational periods, are viewed nearly pole-on or most likely have
small peak-to-peak rotational amplitudes.  The upper limits for the
objects short-term rotational variability as shown in Table 2 were
determined through a monte carlo simulation.  The monte carlo
simulation determined the lowest possible amplitude that would be seen
in the data from the time sampling and variance of the photometry as
well as the errors on the individual points.

Ortiz et al. (2006) reported a possible $0.04\pm 0.02$ photometric
range for (90482) Orcus 2004 DW and a period near 10 hours.  I do not
confirm this result here.  Ortiz et al. (2006) also reported a
marginal $0.08\pm 0.03$ photometric range for (55565) 2002 AW$_{197}$
with no one clear best period.  I can not confirm this result and find
that for 2002 AW$_{197}$ the rotational variability appears
significantly less than 0.08 magnitudes.

Some of the KBOs in this sample appear to have variability which is
just below the threshold of the data detection and thus no significant
period could be obtained with the current data.  In particular 2001
UQ$_{18}$ appears to have a light curve with a significant amplitude
above 0.1 magnitudes but the data is sparser for this object than most
the others and thus no significant period is found.  Followup
observations will be required in order to determine if most of these
flat light curve objects do have any significant short-term
variability.

\subsection{Comparisons with Size, Amplitude, Period, and MBAs}

\begin{figure}
\epsscale{0.7}
\centerline{\includegraphics[angle=90,width=\textwidth]{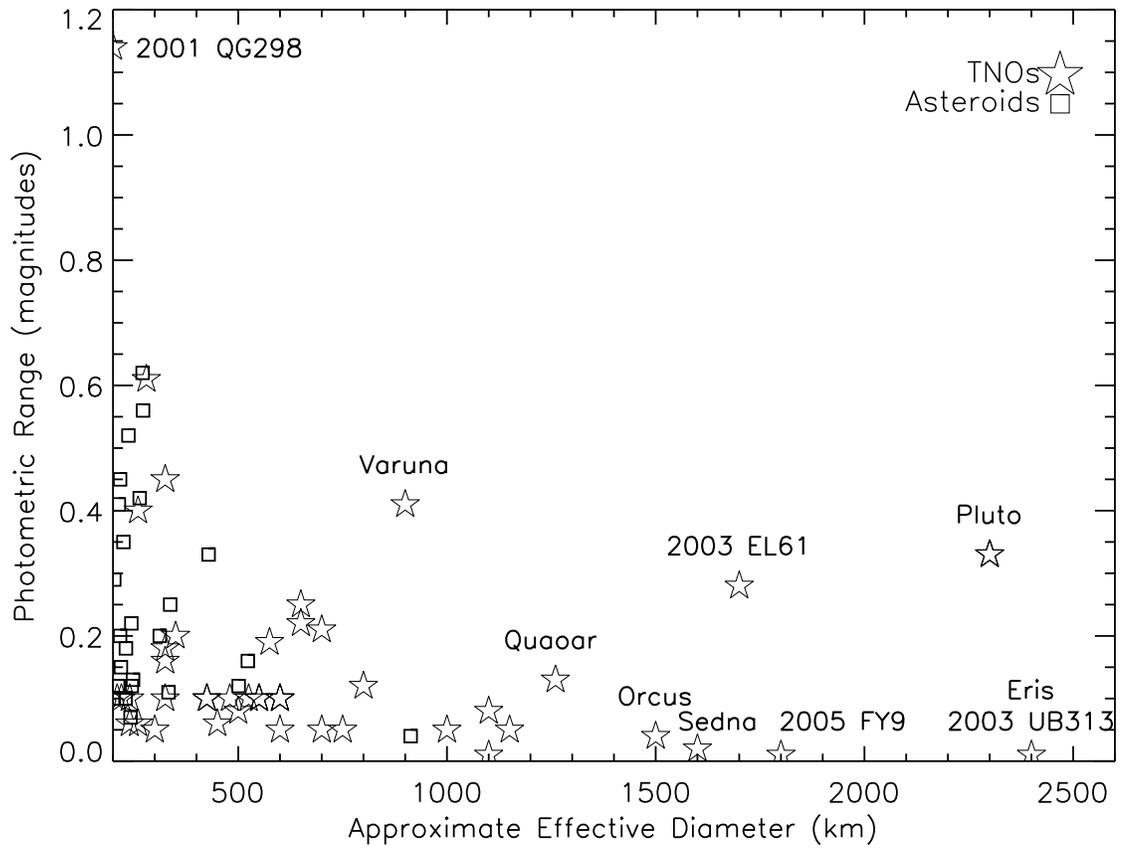}}
\caption{This plot shows the diameter of asteroids and TNOs versus
their light curve amplitudes.  The TNOs sizes if unknown assume they
have moderate albedos of about 10 percent.  For objects with flat
light curves they are plotted at the variation limit found by
observations.}
\label{fig:ampdia} 
\end{figure}

\begin{figure}
\epsscale{0.7}
\centerline{\includegraphics[angle=90,width=\textwidth]{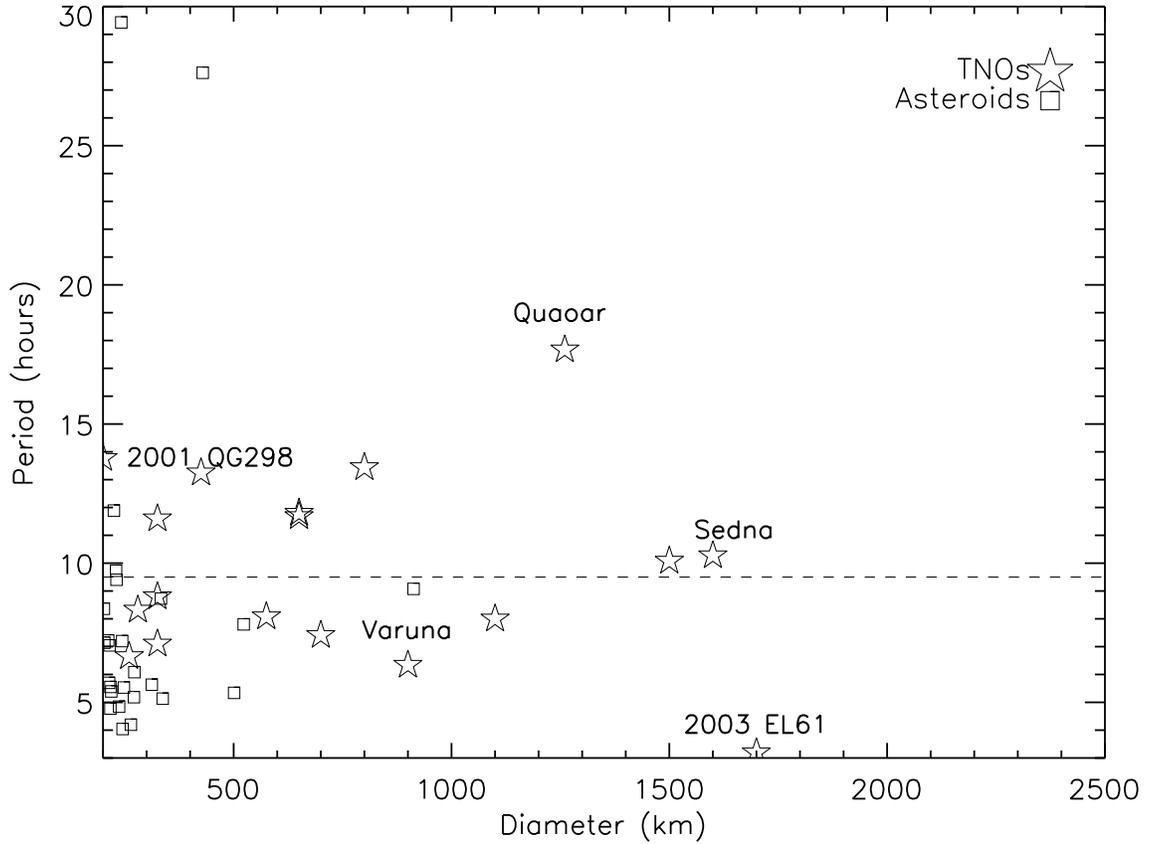}}
\caption{Same as the previous figure except the diameter versus the
light curve period is plotted.  The dashed line is the median of known
TNOs rotation periods ($9.5 \pm 1$ hours) which is significantly above
the median large MBAs rotation periods ($7.0 \pm 1$ hours).  Pluto
falls off the graph in the upper right corner because of its slow
rotation created by the tidal locking to its satellite Charon.}
\label{fig:perdia} 
\end{figure}

In Figures~\ref{fig:ampdia} and \ref{fig:perdia} are plotted the
diameters of the largest TNOs and Main Belt Asteroids (MBAs) versus
rotational amplitude and period, respectively.  Most outliers on
Figure~\ref{fig:ampdia} can easily be explained from the discussion in
section 3.  Varuna, 2003 EL$_{61}$ and the other unmarked TNOs with
photometric ranges above about 0.4 magnitudes are all spinning faster
than about 8 hours.  They are thus likely hydrostatic equilibrium
triaxial Jacobian ellipsoids which are elongated from their rotational
angular momentum (Jewitt and Sheppard 2002; Sheppard and Jewitt 2002;
Rabinowitz et al. 2006).  2001 QG$_{298}$'s large photometric range is
probably because this object is a contact binary indicative of its
longer period and notched shaped light curve (Sheppard and Jewitt
2004).  Pluto's relatively large amplitude light curve is best
explained through its active atmosphere (Spencer et al. 1997).  Like
the MBAs, the photometric amplitudes of the TNOs start to increase
significantly at sizes less than about 300 km in diameter.  The likely
reason is this size range is where the objects are still large enough
to be dominated by self-gravity and are not easily disrupted through
collisions but can still have their angular momentum highly altered
from the collisional process (Farinella et al. 1982; Davis and
Farinella 1997).  Thus this is the region most likely to be populated
by high angular momentum triaxial Jacobian ellipsoids (Farinella et
al. 1992).

From this work Eris (2003 UB$_{313}$) has one of the highest
signal-to-noise time-resolved photometry measurements of any TNO
searched for a rotational period.  There is no obvious rotational
light curve larger than about 0.01 magnitudes in our extensive data
which indicates a very uniform surface, a rotation period of over a
few days or a pole-on viewing geometry.  Carraro et al. (2006) suggest
a possible 0.05 magnitude variability for Eris between nights but this
is not obvious in this data set.  The similar inferred composition and
size of Eris to Pluto suggests these objects should behave very
similar (Brown et al. 2005,2006).  Since Pluto has a relatively
substantial atmosphere at its current position of about 30 AU (Elliot
et al. 2003; Sicardy et al. 2003) it is very likely that Eris has an
active atmosphere when near its perihelion of 38 AU.  At Eris' current
distance of 97 AU its surface thermal temperature should be over 20
degrees colder than when at perihelion.  Like Pluto, Eris' putative
atmosphere near perihelion would likely be composed of N$_{2}$,
CH$_{4}$ or CO which would mostly condense when near aphelion (Spencer
et al. 1997; Hubbard 2003), effectively resurfacing the TNO every few
hundred years.  This is the most likely explanation as to why the
surface of Eris appears so uniform.  This may also be true for 2005
FY$_{9}$ which appears compositionally similar to Pluto (Licandro et
al. 2006) and at 52 AU is about 15 degrees colder than Pluto.

Figure~\ref{fig:perdia} shows that the median rotation period
distribution for TNOs is about $9.5\pm 1$ hours which is marginally
larger than for similarly sized main belt asteroids ($7.0\pm 1$
hours)(Sheppard and Jewitt 2002; and Lacerda and Luu 2006).  If
confirmed, the likely reason for this difference are the collisional
histories of each reservoir as well as the objects compositions.

\section{Phase Curve Results}

\begin{figure}
\epsscale{0.7}
\centerline{\includegraphics[angle=90,width=\textwidth]{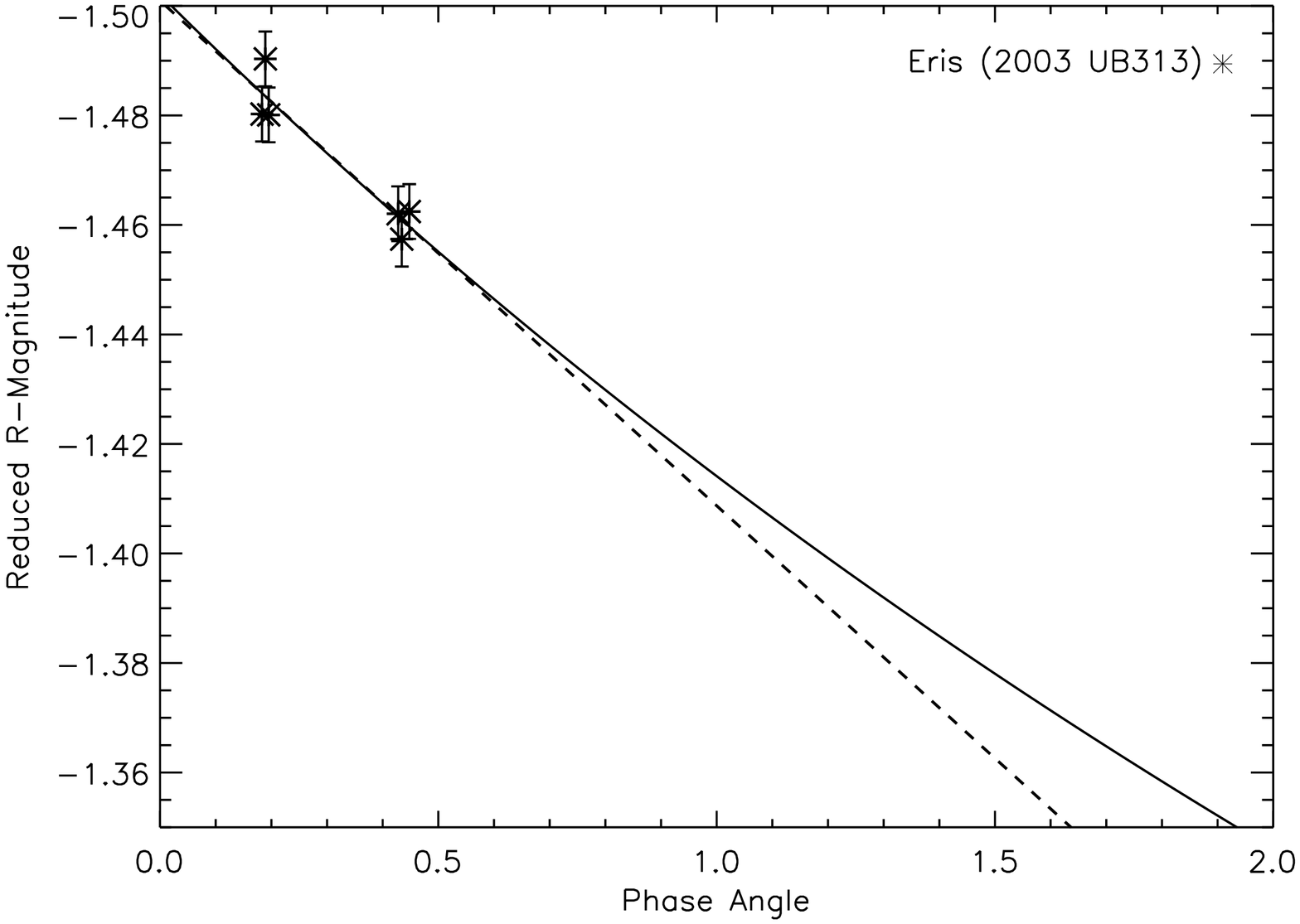}}
\caption{The phase curve for Eris (2003 UB$_{313}$).  The dashed line
is the linear fit to the data while the solid line uses the Bowell et
al. (1989) H-G scattering formalism.  In order to create only a few
points with small error bars, the data has been averaged for each observing night.}
\label{fig:phaseub} 
\end{figure}

\begin{figure}
\epsscale{0.7}
\centerline{\includegraphics[angle=90,width=\textwidth]{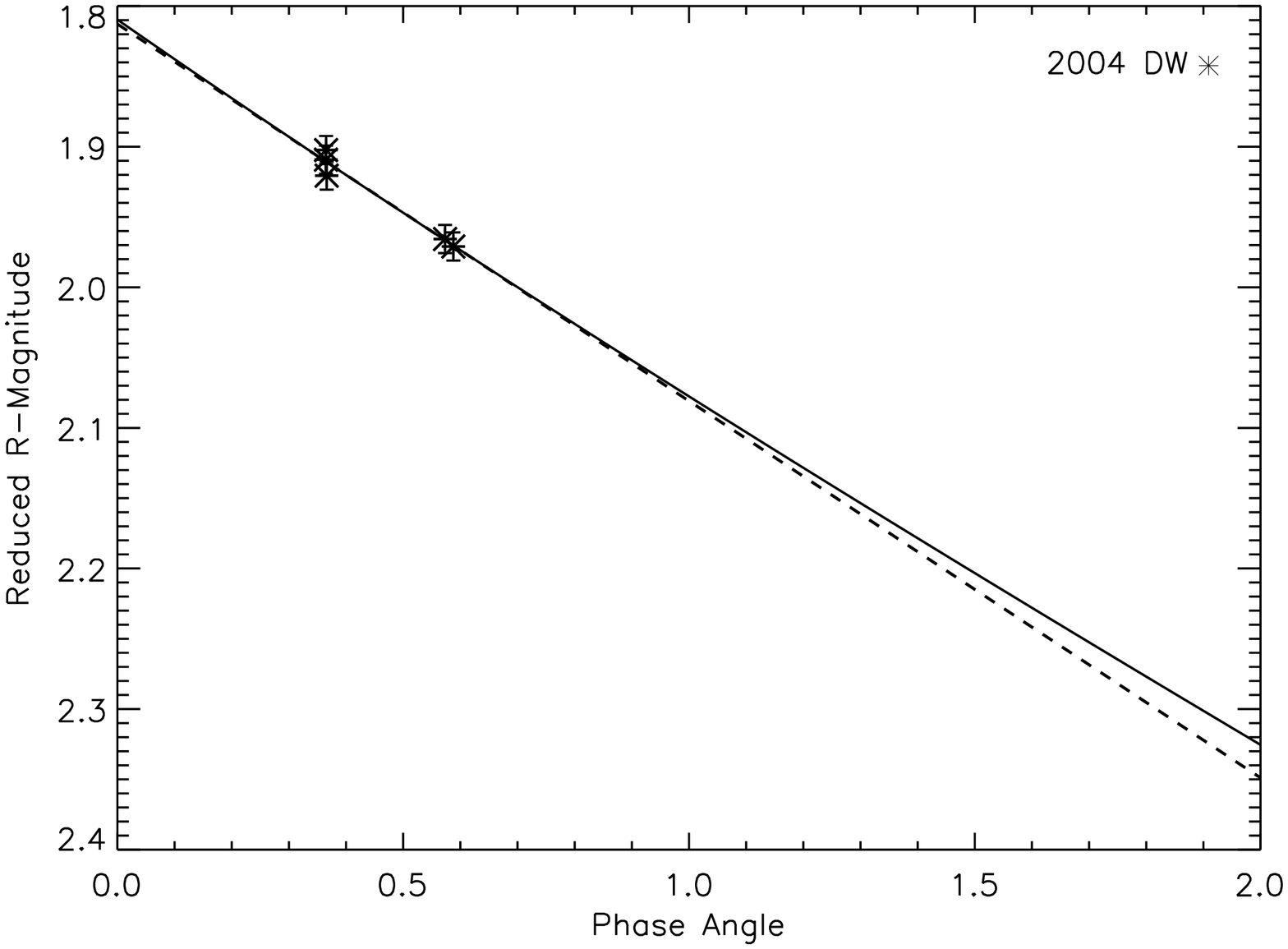}}
\caption{The phase curve for (90482)  Orcus 2004 DW.  The dashed line is
the linear fit to the data while the solid line uses the Bowell et
al. (1989) H-G scattering formalism.  In order to create only a few
points with small error bars, the data has been averaged for each observing night.}
\label{fig:phasedw} 
\end{figure}

\begin{figure}
\epsscale{0.7}
\centerline{\includegraphics[angle=90,width=\textwidth]{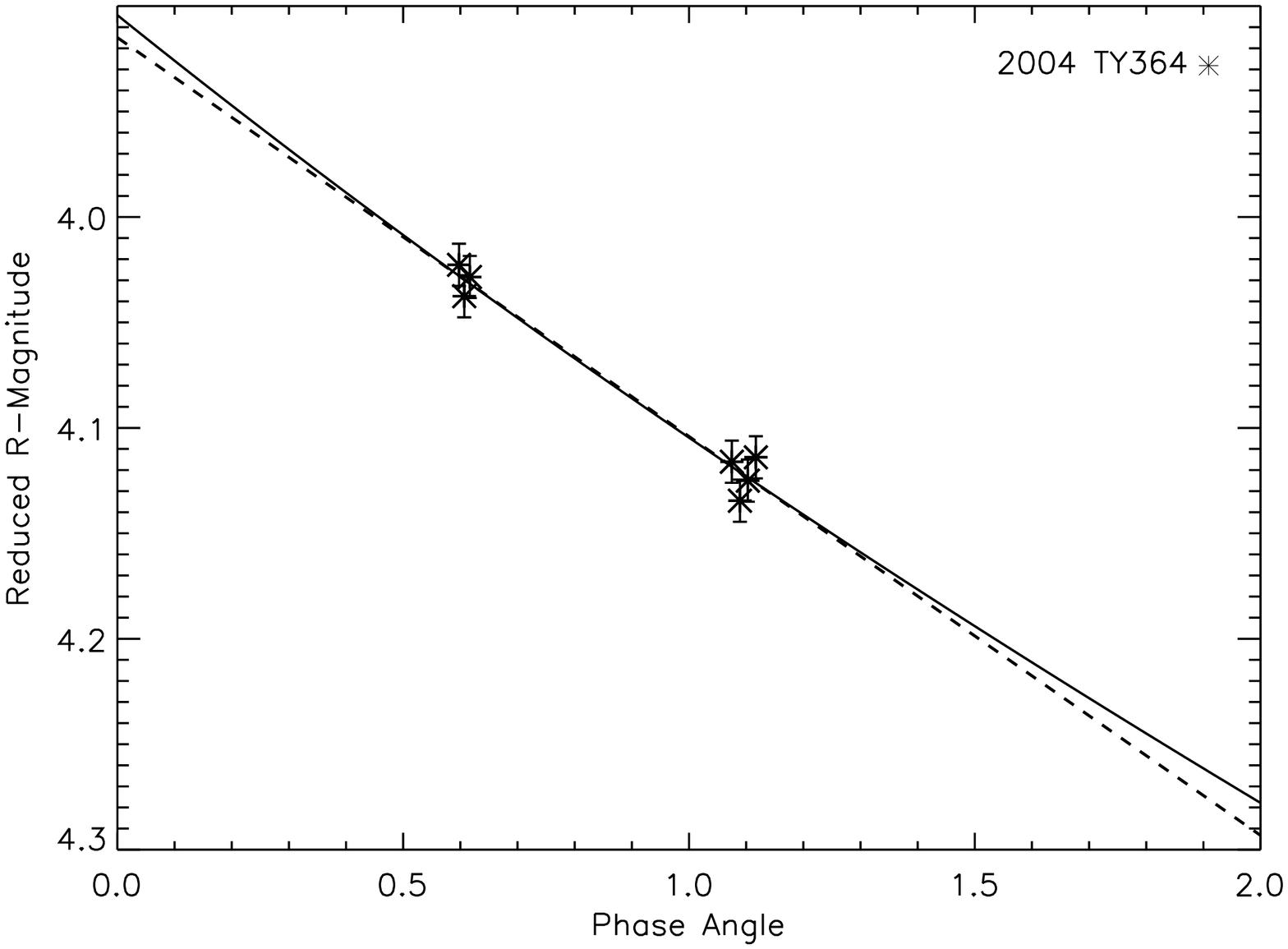}}
\caption{The phase curve for (120348) 2004 TY$_{364}$.  The dashed line is
the linear fit to the data while the solid line uses the Bowell et
al. (1989) H-G scattering formalism.  In order to create only a few
points with small error bars, the data has been averaged for each observing night.}
\label{fig:phasety} 
\end{figure}

\begin{figure}
\epsscale{0.7}
\centerline{\includegraphics[angle=90,width=\textwidth]{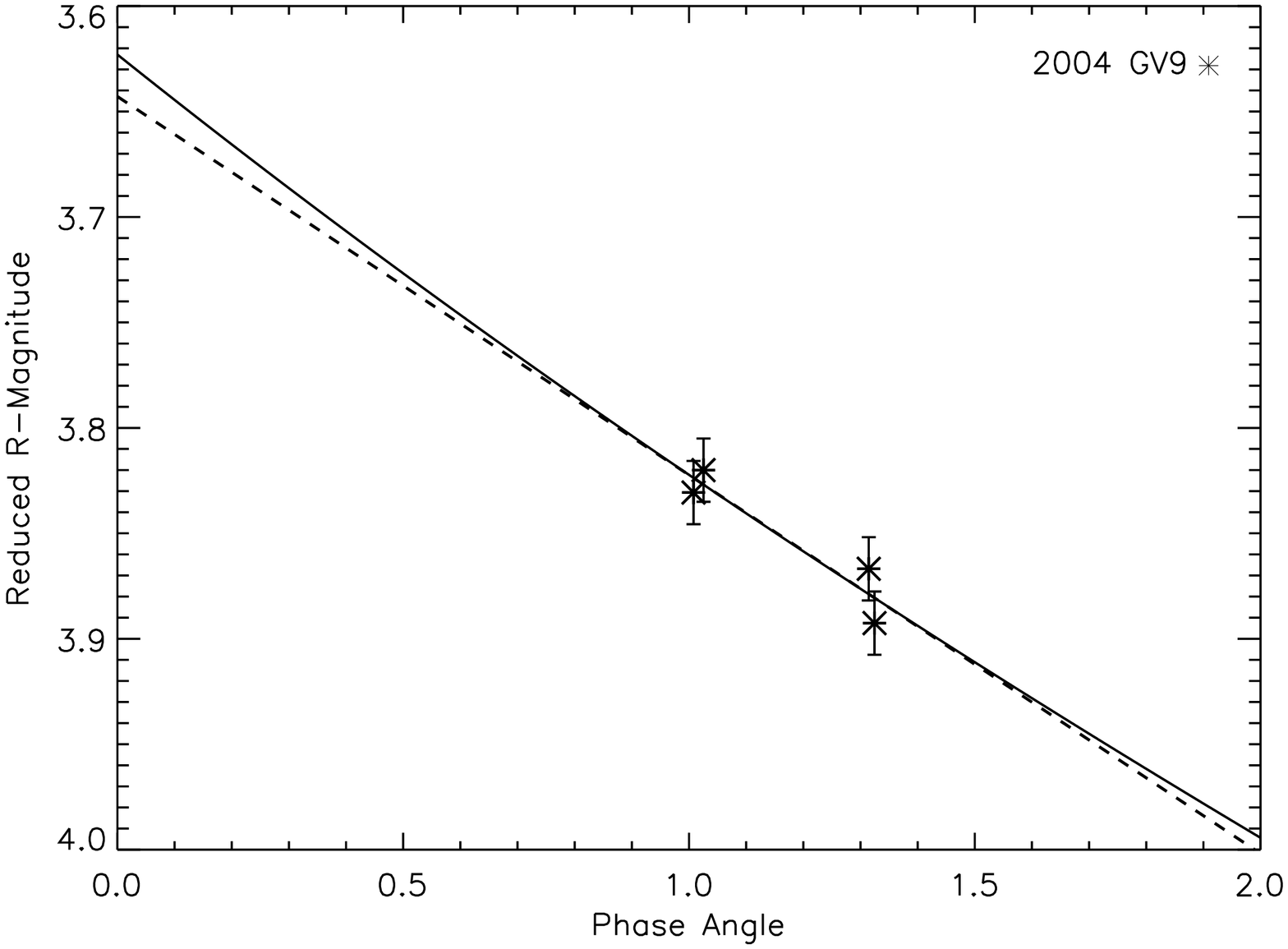}}
\caption{The phase curve for (90568) 2004 GV$_{9}$.   The dashed line is
the linear fit to the data while the solid line uses the Bowell et
al. (1989) H-G scattering formalism.  In order to create only a few
points with small error bars, the data has been averaged for each observing night.}
\label{fig:phasegv} 
\end{figure}

The phase function of an objects surface mostly depends on the albedo,
texture and particle structure of the regolith.  Four of the newly
imaged TNOs (Eris 2003 UB$_{313}$, (120348) 2004 TY$_{364}$, Orcus
2004 DW, and (90568) 2004 GV$_{9}$) were viewed on two separate
telescope observing runs occurring at significantly different phase
angles (Figures~\ref{fig:phaseub} to~\ref{fig:phasegv}).  This allowed
their linear phase functions,
\begin{equation} 
\phi(\alpha) = 10^{-0.4\beta \alpha}
\label{eq:phangle}
\end{equation} 
to be estimated where $\alpha$ is the phase angle in degrees and
$\beta$ is the linear phase coefficient in magnitudes per degree
(Table 3).  The phase angles for TNOs are always less than about 2
degrees as seen from the Earth.  Most atmosphereless bodies show
opposition effects at such small phase angles (Muinonen et al. 2002).
The TNOs appear to have mostly linear phase curves between phase
angles of about 2 and 0.1 degrees (Sheppard and Jewitt 2002,2003;
Rabinowitz et al. 2007).  For phase angles smaller than about 0.1
degrees TNOs may display an opposition spike (Hicks et al. 2005;
Belskaya et al. 2006).

The moderate to large KBOs Orcus, 2004 TY$_{364}$, and 2004 GV$_{9}$
show steep linear R-band phase slopes ($0.18$ to $0.26$ mags per
degree) similar to previous measurements of similarly sized moderate
to large TNOs (Sheppard and Jewitt 2002,2003; Rabinowitz et al. 2007).
In contrast the extremely large dwarf planet Eris (2003 UB$_{313}$)
has a shallower phase slope ($0.09$ mags per degree) more similar to
Charon ($\sim 0.09$ mags/deg; Buie et al. (1997)) and possibly Pluto
($\sim 0.03$ mags/deg; Buratti et al. (2003)).  Empirically lower
phase coefficients between 0.5 and 2 degrees may correspond to bright
icy objects whose surfaces have probably been recently resurfaced such
as Triton, Pluto and Europa (Buie et al. 1997; Buratti et al. 2003;
Rabinowitz et al. 2007).  Thus Eris' low $\beta$ is consistent with it
having an icy surface that has recently been resurfaced.

\begin{figure}
\epsscale{0.7}
\centerline{\includegraphics[angle=90,width=\textwidth]{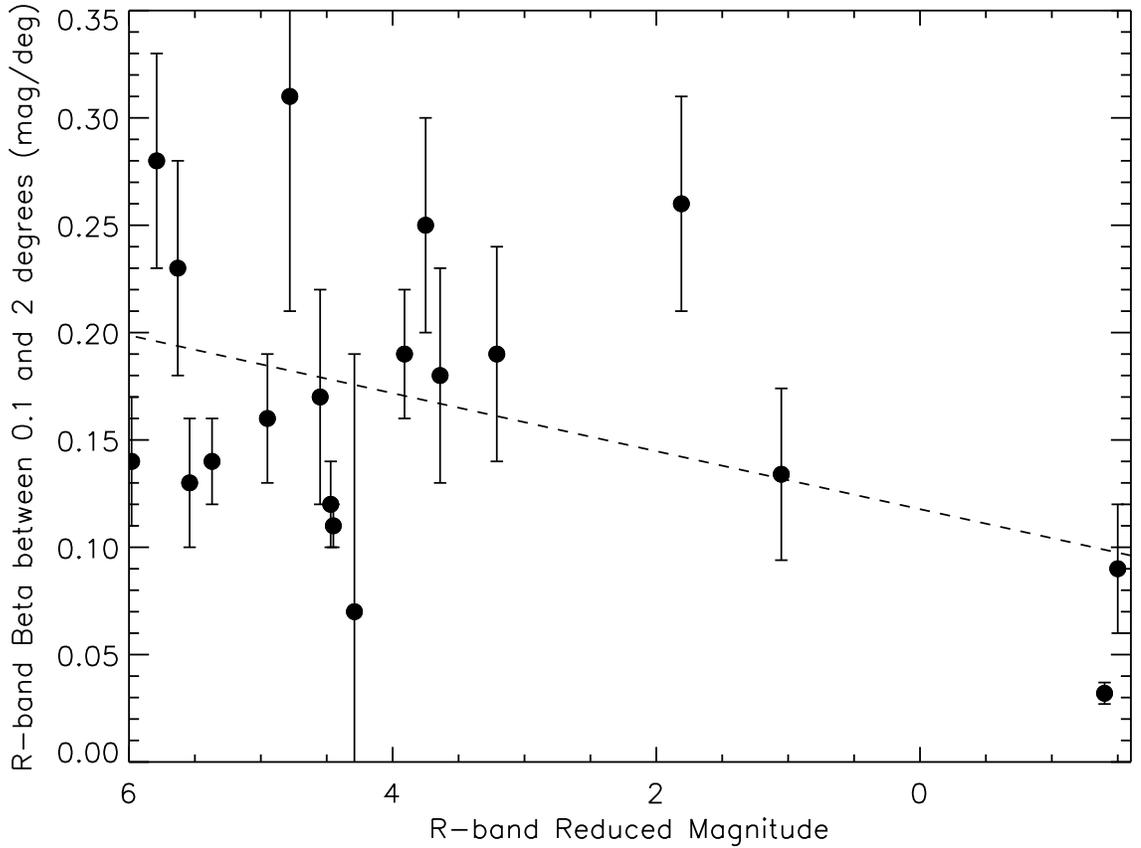}}
\caption{The R-band reduced magnitude versus the R-band linear phase
coefficient $\beta (\alpha < 2$ degrees) for TNOs.  R-band data is
from this work and Sheppard and Jewitt (2002),(2003) as well as Sedna
from Rabinowitz et al. (2007) and Pluto from Buratti et al. (2003).  A
linear fit is shown by the dahsed line.  Larger objects (smaller
reduced magnitudes) may have smaller $\beta$ at the $97 \%$ confidence
level using the Pearson correlation coefficient.}
\label{fig:betaversusHr} 
\end{figure}

\begin{figure}
\epsscale{0.7}
\centerline{\includegraphics[angle=90,width=\textwidth]{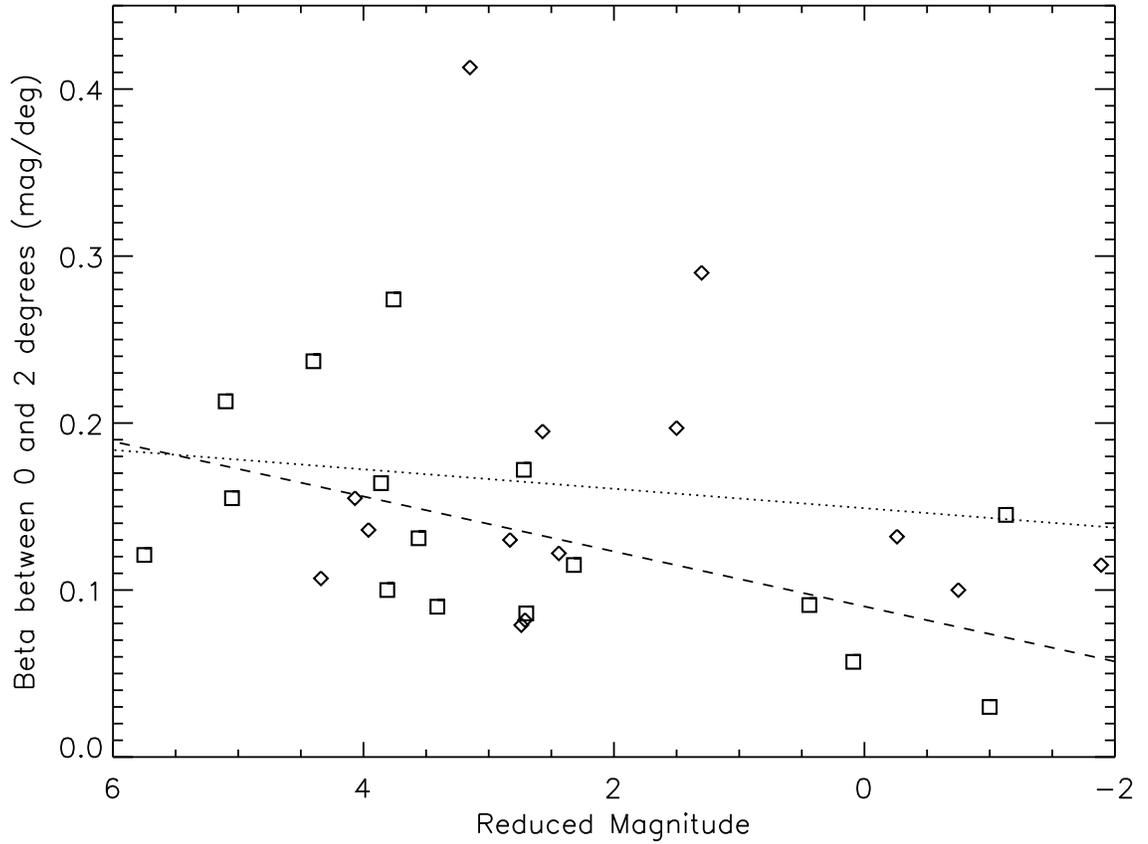}}
\caption{Same as Figure~\ref{fig:betaversusHr} except for the V-band
(squares) and I-band (diamonds). Pluto and Charon data are from Buie
et al. (1997) and the other data are from Rabinowitz et al. (2007).
Error bars are usually less than 0.04 mags/deg.  The V-band data shows
a similar correlation ($97 \%$ confidence, dashed line) as found for
the R-band data in Figure~\ref{fig:betaversusHr}, that is larger
objects may have smaller $\beta$.  There is no correlation found using
the I-band data (dotted line).}
\label{fig:betaversusHvi} 
\end{figure}

\begin{figure}
\epsscale{0.7}
\centerline{\includegraphics[angle=90,width=\textwidth]{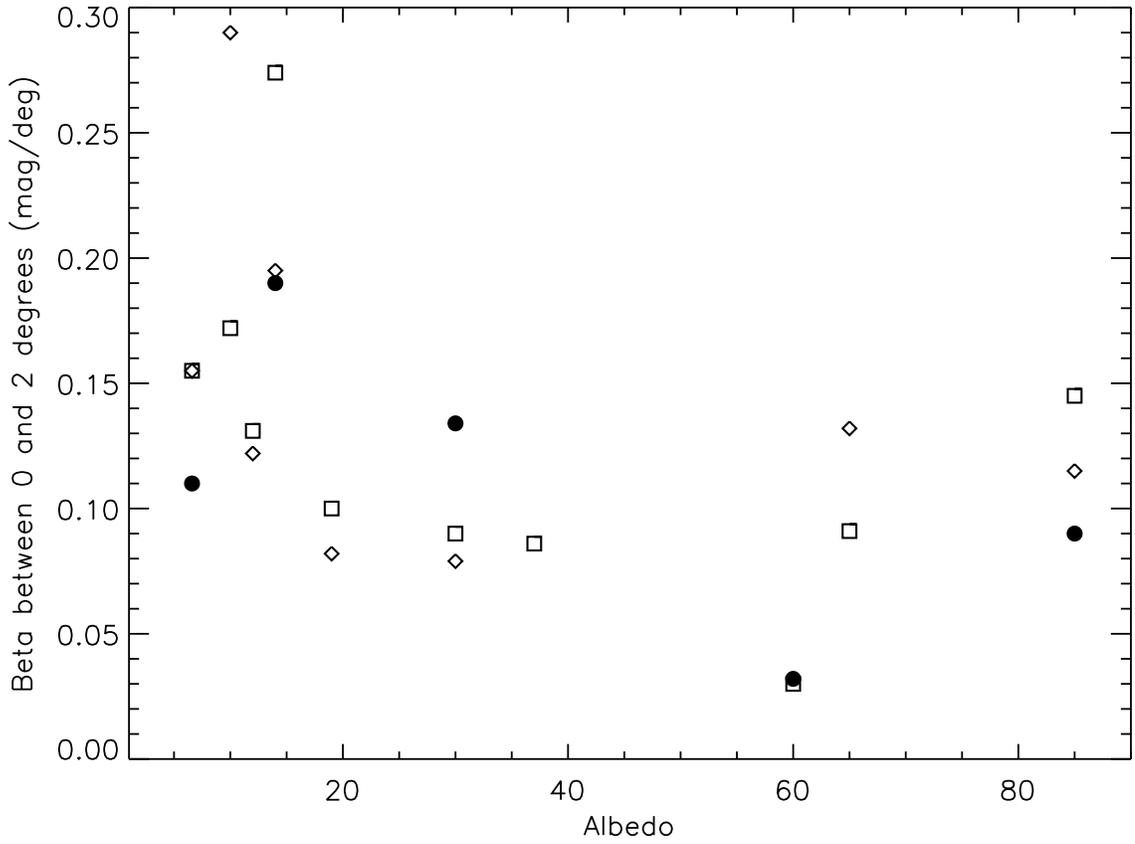}}
\caption{Same as Figures~\ref{fig:betaversusHr}
and~\ref{fig:betaversusHvi} except is the albedo versus linear phase
coefficient for TNOs.  Filled circles are R-band data, squares are
V-band and diamonds are I-band data.  Albedos are from Cruikshank et
al. (2006).}
\label{fig:betaversusalbedo} 
\end{figure}

\begin{figure}
\epsscale{0.7}
\centerline{\includegraphics[angle=90,width=\textwidth]{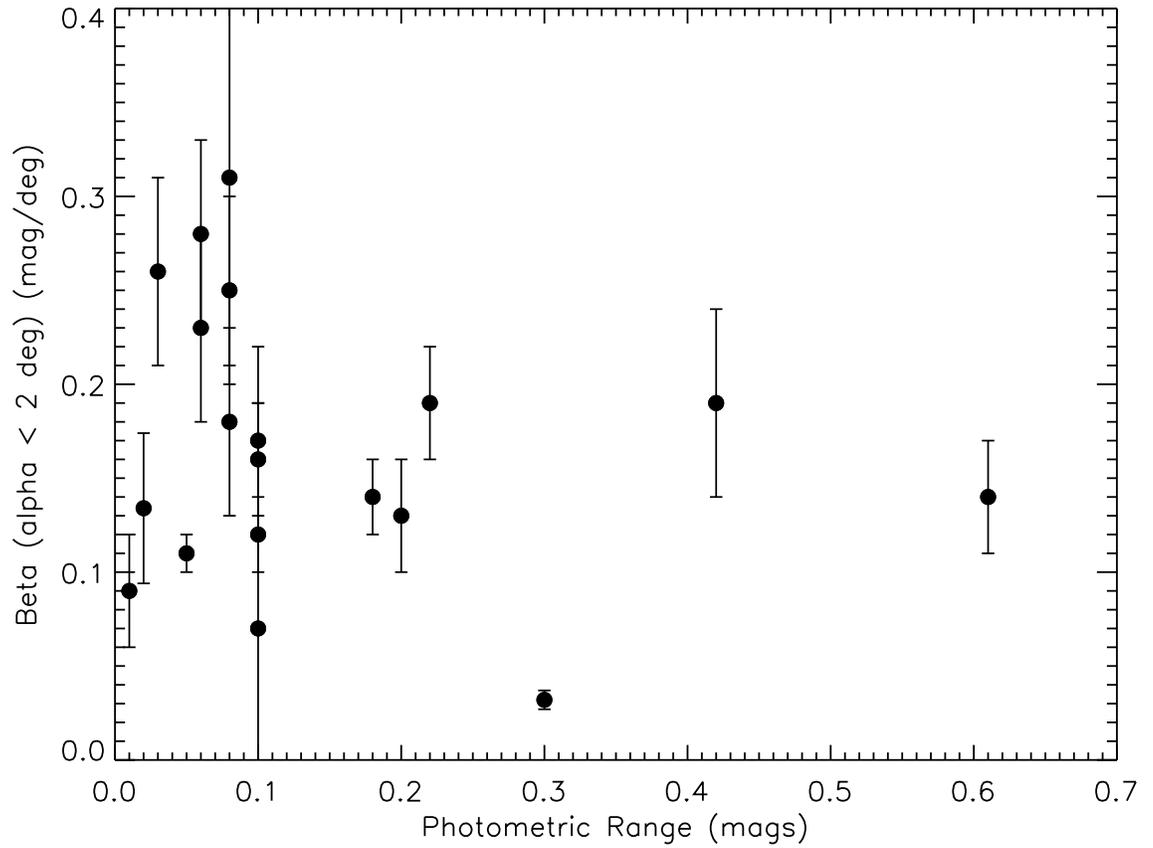}}
\caption{Same as Figure~\ref{fig:betaversusHr} except is the light
curve amplitude versus the linear phase coefficient for TNOs.  TNOs
with no measured rotational variability are plotted with their
possible amplitude upper limits. No significant correlation is found.}
\label{fig:betaversusamp} 
\end{figure}

\begin{figure}
\epsscale{0.7}
\centerline{\includegraphics[angle=90,width=\textwidth]{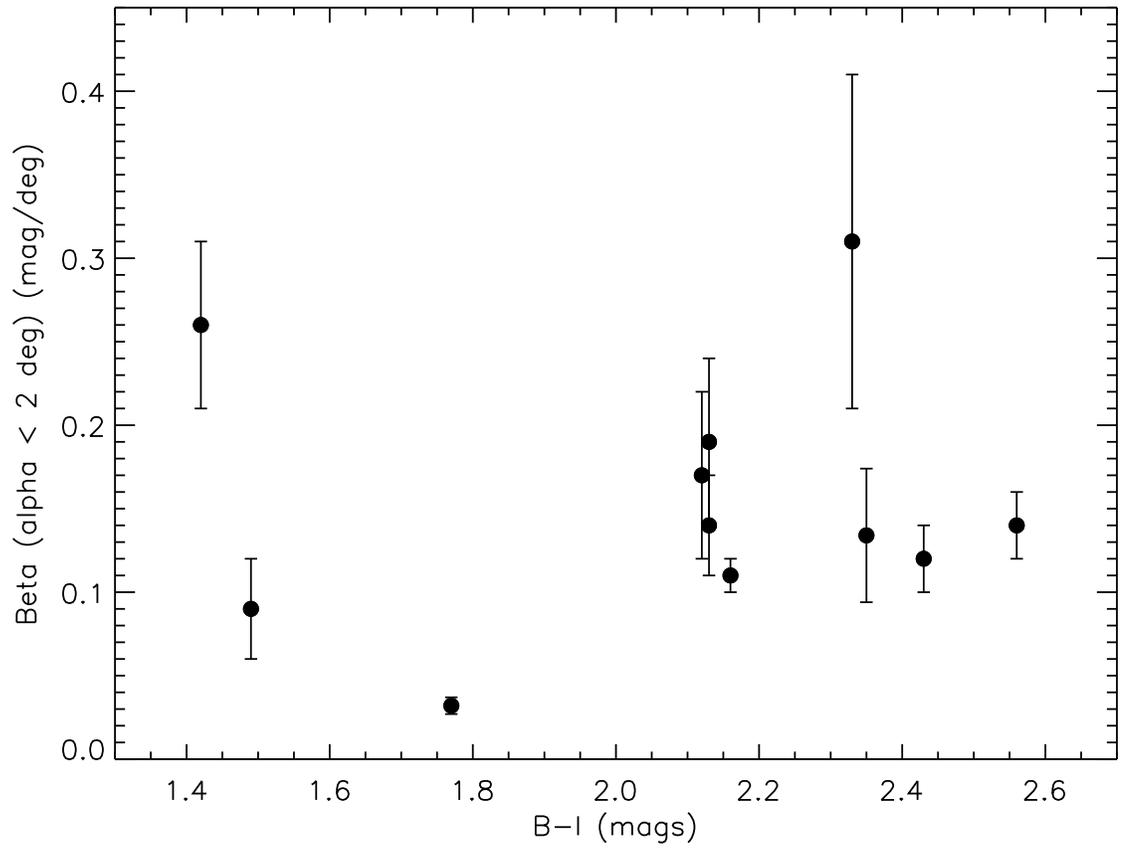}}
\caption{Same as Figure~\ref{fig:betaversusHr} except is the B-I broad
band colors versus the linear phase coefficient for TNOs.  Colors are
from Barucci et al. (2005).  No significant correlation is found.}
\label{fig:betaversuscolor} 
\end{figure}

In Figures~\ref{fig:betaversusHr} to~\ref{fig:betaversuscolor} are
plotted the linear phase coefficients found for several TNOs versus
several different parameters (reduced magnitude, albedo, rotational
photometric amplitude and $B-I$ broad band color).  Table 4 shows the
significance of any correlations.  Based on only a few large objects
it appears that the larger TNOs may have lower $\beta$ values.  This
is true for the R-band and V-band data at the $97 \%$ confidence level
but interestingly using data from Rabinowitz et al. (2007) no
correlation is seen in the I-band (Table 4).  Thus further
measurements are needed to determine if there is a significantly
strong correlation between the size and phase function of TNOs.
Further, it may be that the albedos are anti-correlated with $\beta$,
but since we have such a small number of albedos known the statistics
don't give a good confidence in this correlation.  If confirmed with
additional observations, these correlations may be an indication that
larger TNOs surfaces are less susceptible to phase angle opposition
effects at optical wavelengths.  This could be because the larger TNOs
have different surface properties from smaller TNOs due to active
atmospheres, stronger self-gravity or different surface layers from
possible differentiation.

\subsection{Absolute Magnitudes}

From the linear phase coefficient the reduced magnitude, $m_{R}(1,1,0)
= m_{R} - 5\mbox{log}(R\Delta)$ or absolute magnitude $H$ (Bowell et
al. 1989), which is the magnitude of an object if it could be observed
at heliocentric and geocentric distances of 1 AU and a phase angle of
0 degrees, can be estimated (see Sheppard and Jewitt 2002 for further
details).  The results for $m_{R}(1,1,0)$ and $H$ are found to be
consistent to within a couple hundreths of a magnitude (Table 3 and
Figures~\ref{fig:phaseub} to~\ref{fig:phasegv}).  It is found that the
R-band empirically determined absolute magnitudes of individual TNOs
appears to be several tenths of a magnitude different than what is
given by the Minor Planet Center (Table 3).  This is likely because
the MPC assumes a generic phase function and color for all TNOs while
these two physical properties appear to be significantly different for
individual KBOs (Jewitt and Luu 1998).  The work by Romanishin and
Tegler (2005) attempts to determine various absolute magnitudes of
TNOs by using main belt asteroid type phase curves which are not
appropriate for TNOs (Sheppard and Jewitt 2002).

\section{Summary}

Ten large trans-Neptunian objects were observed in the R-band to
determine photometric variability on times scales of hours, days and
months.

1) Three of the TNOs show obvious short-term photometric variability
   which is taken to correspond to their rotational states.

\begin{itemize}

\item (120348) 2004 TY$_{364}$ shows a double-peaked period of 11.7
hours and if single-peaked is 5.85 hours.  The peak-to-peak amplitude
of the light curve is $0.22\pm 0.02$ mags.

\item (84922) 2003 VS$_{2}$ has a well defined double-peaked period of
  7.41 hours with a peak-to-peak amplitude of $0.21\pm 0.02$ mags.  If
  the light curve is from elongation than 2003 VS$_{2}$'s $a/b$ axis
  ratio is at least 1.2 and the $a/c$ axis ratio is about 1.9.
  Assuming 2003 VS$_{2}$ is elongated from its high angular momentum
  and is a strengthless rubble pile it would have a minimum density of
  about 720 kg m$^{-3}$.

\item (126154) 2001 YH$_{140}$ has a single-peaked period of about 13.25
hours with a photometric range of $0.21\pm 0.04$ mags.

\end{itemize}

2) Seven of the TNOs show no short-term photometric variability within
   the measurement uncertainties.

\begin{itemize}

\item Photometric measurements of the large TNOs (90482) Orcus and
  (55565) 2002 AW$_{197}$ showed no variability within or
  uncertainties. Thus these measurements do not confirm possible small
  photometric variability found for these TNOs by Ortiz et al. (2006).

\item No short-term photometric variability was found for (136199)
  Eris 2003 UB$_{313}$ to about the 0.01 magnitude level.  This high
  signal to noise photometry suggests Eris is nearly spherical with a
  very uniform surface.  Such a nearly uniform surface may be
  explained by an atmosphere which is frozen onto the surface of Eris
  when near aphelion.  The atmosphere, like Pluto's, may become active
  when near perihelion effectively resurfacing Eris every few hundred
  years.  The Methane rich TNO 2005 FY$_{9}$ may also be in a similar
  situation.

\end{itemize}

3) Four of the TNOs were observed over significantly different phase
   angles allowing their long term photometric variability to be
   measured between phase angles of 0.1 and 1.5 degrees.

\begin{itemize}

\item TNOs Orcus, 2004 TY$_{364}$ and 2004 GV$_{9}$ show steep linear
R-band phase slopes between 0.18 and 0.26 mags/degree.

\item Eris 2003 UB$_{313}$ shows a shallower R-band phase slope of 0.09
mags/degree.  This is consistent with Eris having a high albedo, icy
surface which may have recently been resurfaced.

\item At the $97\%$ confidence level the largest TNOs have shallower
R-band linear phase slopes compared to smaller TNOs.  The largest TNOs
surfaces may differ from the smaller TNOs because of their more
volatile ice inventory, increased self-gravity, active atmospheres,
differentiation process or collisional history.

\end{itemize}

3) The absolute magnitudes determined for several TNOs through
   measuring their phase curves show a difference of several tenths of
   a magnitude from the Minor Planet Center values.

\begin{itemize}

\item The values found for the reduced magnitude, $m_{R}(1,1,0)$, and absolute
magnitude, $H$, are similar to within a few hundreths of a magnitude
for most TNOs.

\end{itemize}

\section*{Acknowledgments}

Support for this work was provided by NASA through Hubble Fellowship
grant \# HF-01178.01-A awarded by the Space Telescope Science
Institute, which is operated by the Association of Universities for
Research in Astronomy, Inc., for NASA, under contract NAS 5-26555.

\newpage

%
%
%
%



\begin{center}

\end{center}


\newpage

%



\begin{center}
\begin{deluxetable}{lllcccc}
\tablenum{3}
\tablewidth{6.0 in}
\tablecaption{Phase Function Data for KBOs}
\tablecolumns{5}
\tablehead{
\multicolumn{3}{l}{Name} & \colhead{$m_{R}(1,1,0)$\tablenotemark{a}} & \colhead{$H$\tablenotemark{b}} & \colhead{MPC\tablenotemark{c}}  & \colhead{$\beta (\alpha <2\arcdeg )$\tablenotemark{d}} \\ \colhead{} & \colhead{} & \colhead{} & \colhead{(mag)} & \colhead{(mag)} & \colhead{(mag)} & \colhead{(mag/deg)}}
\startdata
 (136199) & Eris                    & 2003 UB$_{313}$ & $-1.50\pm0.02$  & $-1.50\pm 0.02$  & $-1.65$  &  $0.09\pm 0.03$  \nl
 (90482)  & Orcus                   & 2004 DW         & $1.81\pm0.05$  & $1.81\pm 0.05$   & $1.93$  &  $0.26\pm 0.05$  \nl
 (90568)  &                         & 2004 GV$_{9}$   & $3.64\pm0.06$  & $3.62\pm 0.06$   & $3.5$  &  $0.18\pm 0.06$  \nl
 (120348) &                         & 2004 TY$_{364}$ & $3.91\pm0.03$  & $3.90\pm 0.03$   & $4.0$  &  $0.19\pm 0.03$  \nl
\enddata
\tablenotetext{a}{The R-band reduced magnitude determined from the linear phase coefficient found in this work.}
\tablenotetext{b}{The R-band absolute magnitude determined as described in Bowell et al. (1989).}
\tablenotetext{c}{The R-band absolute magnitude from the Minor Planet Center converted from the V-band as is shown in Table 2 to the R-band using the known colors of the objects: V-R$=0.45$ for Eris (Brown et al. 2005), V-R$=0.37$ for Orcus (de Bergh et al. 2005), and a nominal value of V-R$=0.5$ for 90568 and 120348 since these objects don't have known V-R colors.}
\tablenotetext{d}{$\beta (\alpha < 2\arcdeg)$ is the phase coefficient in magnitudes per degree at phase angles $< 2\arcdeg$.}
\end{deluxetable}
\end{center}


\newpage

%



\begin{center}
\begin{deluxetable}{lccc}
\tablenum{4}
\tablewidth{5 in}
\tablecaption{Phase Function Correlations}
\tablecolumns{4}
\tablehead{
\colhead{$\beta$ vs.\tablenotemark{a}} & \colhead{$r_{corr}$\tablenotemark{b}} & \colhead{N\tablenotemark{c}} & \colhead{Sig\tablenotemark{d}} \\ \colhead{} & \colhead{} & \colhead{} & \colhead{}}
\startdata
$m_{R}(1,1,0)$  & 0.50   & 19  & $97\%$ \nl
$m_{V}(1,1,0)$  & 0.54   & 16  & $97\%$ \nl
$m_{I}(1,1,0)$  & 0.12   & 14  & $<60\%$ \nl
$p_{R}$      & -0.51  & 5   & $65\%$ \nl
$p_{V}$      & -0.38  & 9   & $70\%$ \nl
$p_{I}$      & -0.27  & 10  & $<60\%$ \nl
$\Delta m$      & -0.21  & 19  & $<60\%$ \nl
$B-I$           & -0.20  & 11  & $<60\%$ \nl
\enddata
\tablenotetext{a}{$\beta$ is the linear phase coefficient in magnitudes per degree at phase angles $< 2\arcdeg$.  In the column are what $\beta$ is compared to in order to see if there is any correlation; $m_{R}(1,1,0)$, $m_{V}(1,1,0)$ and $m_{I}(1,1,0)$ are the reduced mangitudes in the R, V and I-band respectively and are compared to the value of $\beta$ determined at the same wavelength; $p_{R}$, $p_{V}$ and $p_{I}$ are the geometric albedos compared to $\beta$ in the R, V and I-band respectively; $\Delta m$ is the peak-to-peak amplitude of the rotational light curve; and $B-I$ is the color.  The phase curves in the R-band are from this work and Sheppard and Jewitt (2002;2003) while the V and I-band data are from Buie et al. (1997) and Rabinowitz et al. (2007).  The albedo information is from Cruikshank et al. (2006) and the colors from Barucci et al. (2005).}
\tablenotetext{b}{$r_{corr}$ is the Pearson correlation coefficient.}
\tablenotetext{c}{N is the number of TNOs used for the correlation.}
\tablenotetext{d}{Sig is the confidence of significance of the correlation.}
\end{deluxetable}
\end{center}


\newpage

\begin{figure}
\epsscale{0.7}
\centerline{\includegraphics[angle=90,width=\textwidth]{fig01.ps}}
\caption{The Phase Dispersion Minimization (PDM) plot for (120348) 2004
TY$_{364}$.  The best fit single-peaked period is near 5.85 hours.}
\label{fig:pdmty} 
\end{figure}

\clearpage

\clearpage

\clearpage

\clearpage

\clearpage

\clearpage

\clearpage

\clearpage

\clearpage

\clearpage

\clearpage

\clearpage

\clearpage

\clearpage

\clearpage

\clearpage

\clearpage

\clearpage

\clearpage

\clearpage

\clearpage

\clearpage

\clearpage

\clearpage

\clearpage

\clearpage

\clearpage

\clearpage

\clearpage

\clearpage

\end{document}